\documentclass[draftcls,onecolumn,12pt]{IEEEtran}

\usepackage{cite}
\usepackage{amsmath,amssymb,amsfonts}
\usepackage{algorithmic}
\usepackage{graphicx}
\usepackage{textcomp}
\usepackage{xcolor}
\usepackage{subfigure}
\usepackage{caption}
\usepackage{ifpdf}
\usepackage{float}
\usepackage{booktabs}

\ifCLASSINFOpdf
\else
\fi
\usepackage{amsmath}
\usepackage{algorithmic}

\usepackage{array}
\begin{document}
%
\title{Performance of Wireless Optical Communication With Reconfigurable Intelligent Surfaces and Random Obstacles}
%
%
%

\author{Haibo Wang,~\IEEEmembership{Member,~IEEE,}
        Zaichen Zhang,~\IEEEmembership{Senior Member,~IEEE,} Bingcheng Zhu,~\IEEEmembership{Member,~IEEE,} Jian Dang,~\IEEEmembership{Member,~IEEE,} Liang Wu,~\IEEEmembership{Member,~IEEE,}
Lei Wang, Kehan Zhang and~, Yidi Zhang}

\maketitle

\begin{abstract}
It is difficult for free space optical communication to be applied in mobile communication due to the obstruction of obstacles in the environment, which is expected to be solved by reconfigurable intelligent surface technology. The reconfigurable intelligent surface is a new type of digital coding meta-materials, which can reflect, compute and program electromagnetic and optical waves in real time. We purpose a controllable multi-branch wireless optical communication system based on the optical reconfigurable intelligent surface technology. By setting up multiple optical reconfigurable intelligent surface in the environment, multiple artificial channels are built to improve system performance and to reduce the outage probability. Three factors affecting channel coefficients are investigated in this paper, which are beam jitter, jitter of the reconfigurable intelligent surface and the probability of obstruction. Based on the model, we derive the closed-form probability density function of channel coefficients, the asymptotic system's average bit error rate and outage probability for systems with single and multiple branches. It is revealed that the probability density function contains an impulse function, which causes irreducible error rate and outage probability floors. Numerical results indicate that compared with free-space optical communication systems with single direct path, the performance of the multi-branch system is improved and the outage probability is reduced.
\end{abstract}


%
\IEEEpeerreviewmaketitle

\section{Introduction}\label{introduction}

\IEEEPARstart{A}{fter} 2020, the fifth generation of mobile communications (5G) is expected to achieve global commercialization. From the second-generation mobile communication (2G) to 5G, the communication frequency band has been increased from 100 MHz to GHz \cite{lin2019millimeter}\cite{ni2019research}. Higher frequency electromagnetic waves are exploited for more spectrum resources. In order to discover new spectrum resources, research on millimeter waves, terahertz, and optical communications will become important directions\cite{nishimura2019optical}\cite{xiao2019resource}\cite{busari2019terahertz}. For optical wireless communication, free space optical communication (FSO), visible light communication(VLC), short-range near-infrared communication and other technologies have been thoroughly studied and widely applied. However, light waves are easily absorbed by non-transparent obstacles, thus optical communication scenes are usually limited to unobstructed scenarios, i.e. line-of-sight circumstances. In addition, with the increase of communication frequency bands, high-frequency signals such as millimeter waves, terahertz, etc., gradually show similar characteristics to optical signals, such as narrow pulses and easy to be blocked\cite{song2011present}\cite{rebeiz1992millimeter}\cite{siegel2002terahertz}\cite{rappaport2014mobile}. Therefore, a solution is required to reduce the impact of these characteristics on communication quality.\par
Reconfigurable intelligent surface (RIS) is a new type of meta-surface that can programmably modulate the electromagnetic waves passing through it\cite{huang2019reconfigurable}\cite{li2019reconfigurable}\cite{di2019hybrid}\cite{wu2019intelligent}. At present, the RIS structure in the microwave band is mainly composed of an array of digital coding units. The beam incident on each unit can be adjusted to control the intensity, phase, frequency, and polarity of the outgoing beam. In \cite{di2019hybrid}, Boya Di, Hongliang Zhang, etc. proposed to use RIS to implement microwave beamforming, which is equivalent to adjusting the large-scale antenna array of the base station towards multiple nodes in free space. The advantage is to reduce the pressure of the base station and improve the energy utilization efficiency, and the microwave signals that have not been received can be recollected and transmitted.\par
Analogous to the RIS structure in the microwave band, the optical RIS structure needs to achieve the following functions: (1) Reflecting the incident beam; (2) Keeping the information carried by the original beam unchanged or slightly changed; (3) Controlling the intensity, phase, frequency, polarization and other characteristics of the outgoing beam programmably; (4) Adjusting the direction of the outgoing beam precisely to follow the user.\par In the prior technology, spatial light modulator (SLM) and optical micro-electro-mechanical system (MEMS) meet the requirements\cite{vidal2006optical}\cite{armitage1985high}\cite{ross1982two}\cite{ma2003optical}. In 1982, a two-dimensional magneto-optic spatial light modulator was proposed \cite{ross1982two}, which was used to adjust the amplitude, phase, polarization and other parameters of the light passing through it. With the lens group, SLM can reconstruct the light field with low power loss. The SLM is composed of a digital coding unit array, where each unit can programmatically adjust the amplitude and phase of the incident light, and the modulation frequency can reach 100Hz. In \cite{kim2013wireless}, SLM is used for signal modulation in low-speed VLC system. In \cite{zhang2018optical}, SLM is used to convert a single beam at the transmitting end into multiple beams, and generate optical signals that follow multiple mobile users. Optical MEMS is a lens array composed of freely adjustable micro lenses, which can freely adjust the direction of reflected light at each unit. Compared with SLM, it has lower cost and can be mass-produced under the existing technology, but it can not freely control the phase, frequency, and polarization of the outgoing beam.\par
Based on the optical RIS structure, we propose a controllable multi-branch wireless optical communication system with optical RIS in channels, namely optical intelligent channel communication system. By setting multiple optical RIS, namely intelligent channel reconfigurable node (ICRN) in the communication scenario, we can build multiple artificial optical channels, namely intelligent channels. The intelligence of the system is shown in: (1) For mobile users, the transmitter and ICRN cooperate to enable the signal to follow the users and be aimed to the user's receiver center; (2) Multiple controllable channels based on ICRN are built between the base station and users, where the channel path can be adjusted by selecting the ICRN nodes that the path passes through; (3) The physical path of each channel is known by the base station and the channel state information (CSI) can be estimated in real time; (4) According to CSI, the base station can allocate power to each channel for power efficiency optimization. The power allocation coefficient can be adjusted in real time to keep the communication stable.\par
The main purpose of this paper is to analyze the performance of the optical intelligent channel communication system. It is assumed that the beam of each channel has been aimed at the center of the receiver. Since the communication distance is set within 500 meters, the influence of atmospheric turbulence can be ignored\cite{sandalidis2008ber}\cite{garcia2009selection}\cite{uysal2006error}. Without loss of generality, the system is assumed to have an ideal receiver array, implying that the receiver receives all the energy of the incident optical signal. Three factors are mainly analyzed in this system, which are beam jitter, ICRN jitter and probability of obstruction. Beam jitter refers to the light beam vibrating due to the jitter at the transmitting end\cite{sandalidis2008ber}\cite{borah2009pointing}\cite{burks1982high}. ICRN jitter refers to the jitter of the ICRN surface, which results in the deflection of the normal vector of the reflecting surface. The probability of obstruction is a new factor that affects channel fading, since the system is assumed to be in an environment with obstacles. The probability should be a quantity that changes slowly over time and varies for different paths and channel lengths.\par
The contributions of this paper are as follows: \par
1) Based on optical RIS technology, we design an optical intelligent channel communication system. The aim is to propose a solution to realize stable optical communication in an environment with obstacles, and to broaden the application scenarios of optical communication. Different from the FSO diversity transmission, the RIS node in the system physically reflects the optical signal without receiving and forwarding the signal, thus reducing the cost and communication delay.\par
2) Physical modeling is performed on the beam jitter and ICRN jitter in systems with RIS and the probability density function (PDF) of pointing error displacement is derived, which is verified by simulation results. Based on the analysis of pointing error and probability of obstruction, the expressions of PDF of SNR, the average bit error rate (BER) and outage probability of systems with single branch and multi-branches are derived, which are verified by simulation results. \par
3) The system performance gain by increasing the number of channels is analyzed, which reveals that increasing the number of intelligent channels with ICRN can improve system performance and reduce outage probability. However, the performance gain by adding an intelligent channel decreases as the number of channels increases.
4) We propose an optimization scheme for power allocation to multiple intelligent channels at high SNR.\par
Other sections of this paper is as follows, Section \ref{model} describes our system model and derives the closed-form PDF of the channel fading. In this model, three new elements are investigated, which are the pointing error when there exists a reflective surface in the optical path, the jitter of the reflective surface and the probability of obstruction in the channel. In Section \ref{analysis}, we derive the expressions of asymptotic BER and outage probability of the systems with single branch and multi-branch. Section \ref{discussion} discusses the performance gain for increasing the number of channels and purposes an optimized power allocation scheme for multi-branch at high SNR. Section \ref{simulation} presents some numerical results, and Section \ref{conclusion} makes several important conclusions.

\section{System Model}\label{model}
\begin{figure}
\centering
\includegraphics[width=1\textwidth]{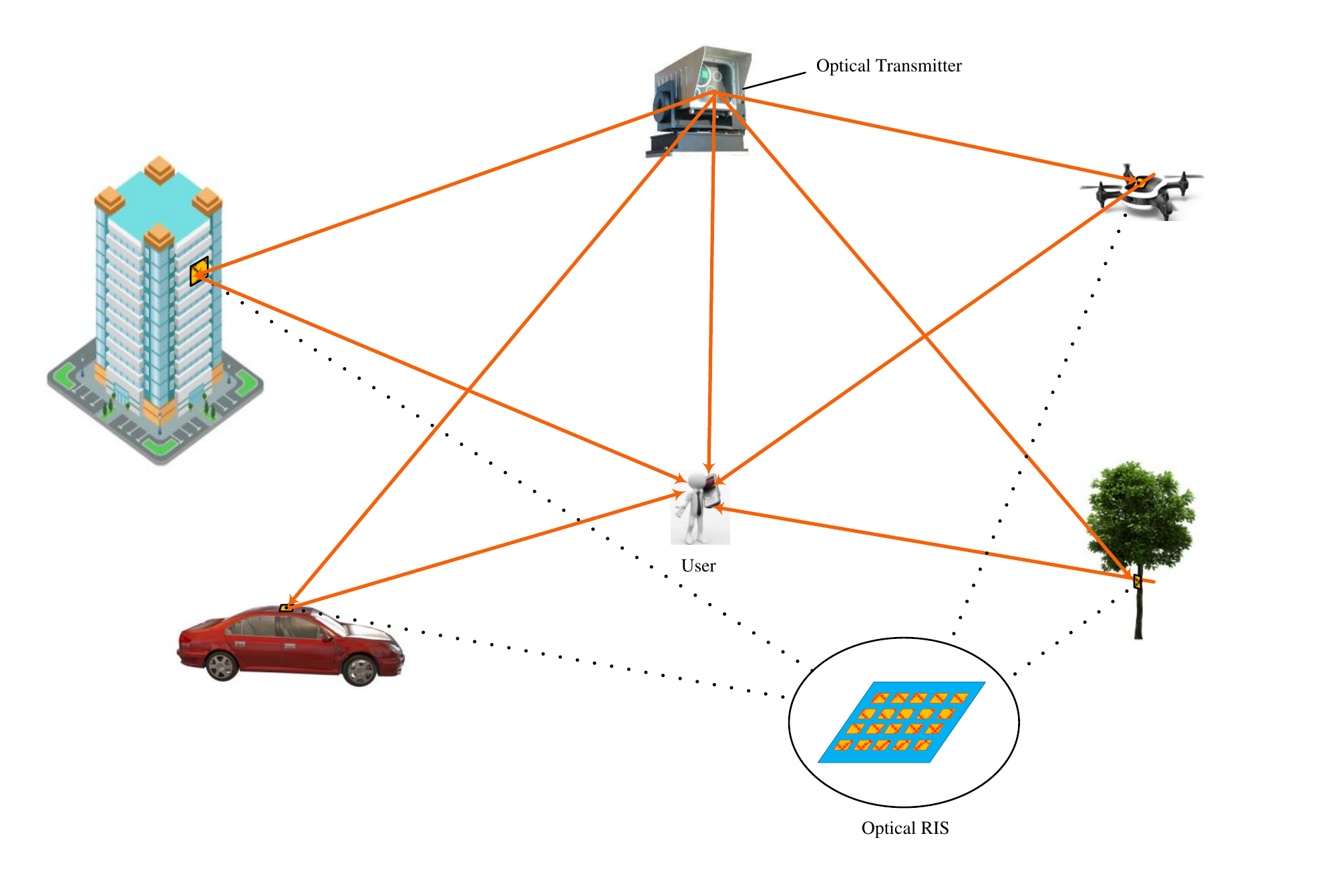}
\caption{optical intelligent channel communication system.}
\label {fig.1}
\end{figure}

As shown in Fig. \ref{fig.1}, in the optical intelligent channel communication system, multiple ICRN nodes are set between the light source and the receiver to build multiple controllable channels, namely intelligent channels. Each ICRN node can deflect the beam without changing the signal's amplitude and phase. Whether ICRN is implemented using optical RIS such as SLM or MEMS, it can be modeled as a mirror that conforms to the law of reflection. Therefore, the jitter of ICRN can be described by the vibration of the mirror's normal vector. In this system, we make the following assumption.\par  AS1) With the cooperation of the transmitter and ICRN, all the beams have been precisely aimed at the center of the receiver.\par AS2) The receiver is ideal. That is, the receiver receives all the energy of the incident optical signal from all directions.\par AS3) $M$ ICRNs are employed in free space. The transmitter transmits signals to all ICRNs simultaneously, and each ICRN directly reflects the signals to the receiver. \par Therefore, there are $M$ intelligent channels in the space and the received signal power $s$ can be presented as
\begin{equation} \label{2-1}
\begin{split}
s=\sum_{k=0}^{M-1}h_ks_{k}+n
\end{split}
\end{equation}
where $s_{k}$ is the signal intensity assigned to $k$th channel, $h_k$ is the channel fading of the $k$th channel and $n$ is the zero-mean Gaussian white noise from the receiver with variance of $\sigma_{n}^2$. In this system, we utilize intensity direct detection (IM/DD) with on-off keying (OOK). The data bits are directly modulated onto the intensity of the optical beam by the transmitter. $s_{k}$ is either $0$ or $2\alpha_kP_t$, where $P_t$ is the average power of the total transmitted signal, $\alpha_k$ is the power allocation coefficient of $k$th channel. Since the system scenario does not involve long-distance communication (above 500 meters), atmospheric noise can be disregarded. In this system, three factors that affect channel conditions are analyzed, which are the pointing error caused by beam jitter and ICRN jitter and the probability of occlusion.

\subsection{Pointing Error}
\begin{figure*}
\centering
\includegraphics[width=1\textwidth]{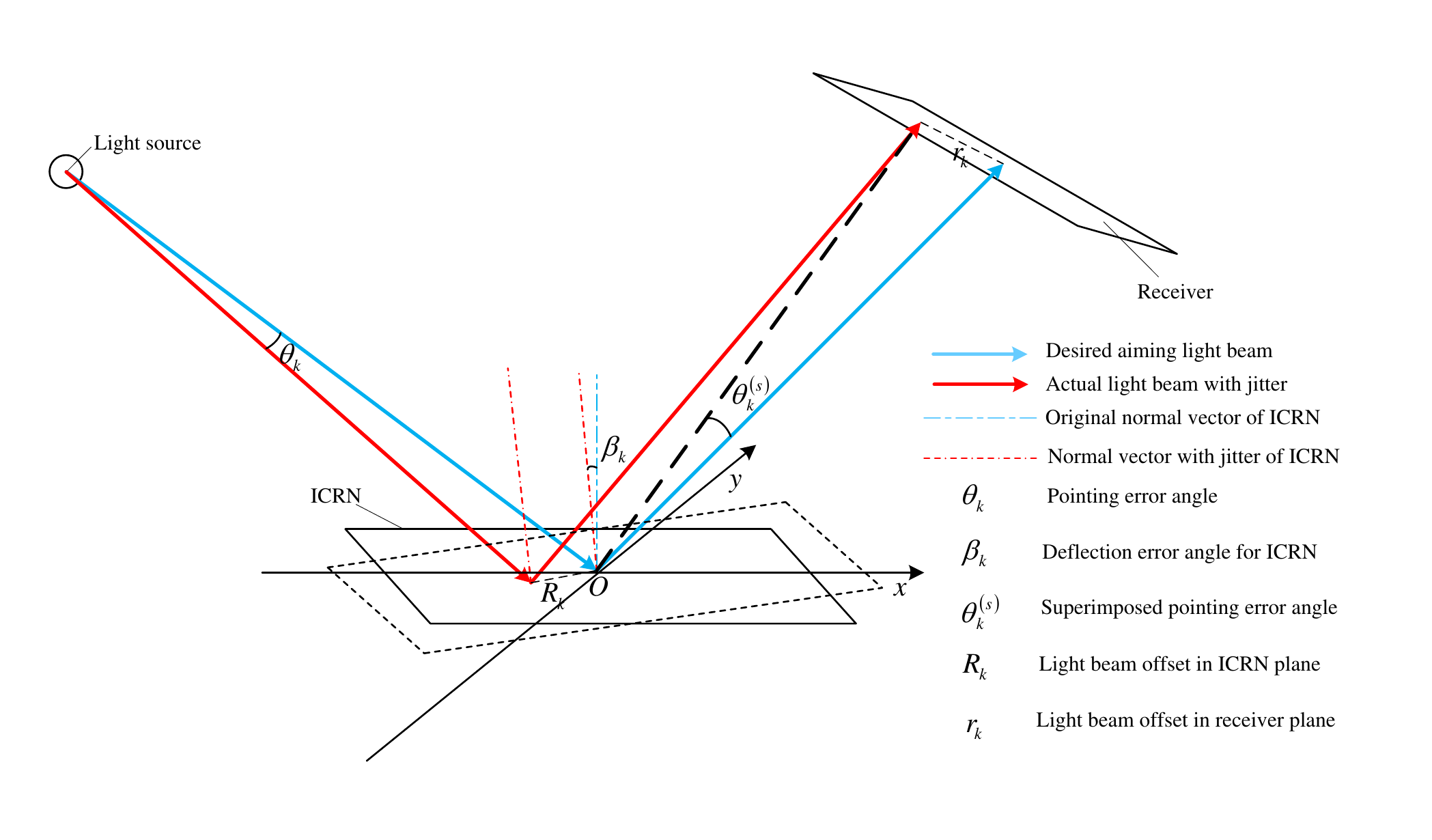}
\caption{Diagram of beam jitter and ICRN jitter in the optical intelligent channel system with single branch.}
\label {fig.2}
\end{figure*}
Due to the mechanical jitter at the transmitter and ICRN, even if the beam has been aimed at the center of the receiver, it will still randomly vibrate within a certain range\cite{sandalidis2008ber}\cite{orzechowski2008optimal}\cite{arancibia2005adaptive}. In this section, we derive a new model for pointing error caused by beam jitter and ICRN jitter. Fig. \ref{fig.2} shows the diagram of beam jitter and ICRN jitter in the optical intelligent channel system with single branch. As shown in Fig. \ref{fig.2}, in the intelligent channel consisting of a transmitter, an ICRN and a receiver, the pointing error angle $\theta_k$ is the angle between the desired aiming light beam and actual emitted light beam with a jitter, which describes the beam jitter and the deflection error angle $\beta_k$ is is the angle between the ICRN original normal vector and ICRN actual normal vector with jitter, which describes the ICRN jitter. The desired aiming light is aimed at the receiver center and is perpendicular to the receiver plane. From Fig. \ref{fig.2}, we can observe that both $\theta_k$ and $\beta_k$ cause the displacement $R_k$ from the receiver center to actual receiving light spot. In the ICRN plane, a two-dimensional Cartesian coordinate system is established, where the coordinate origin is set where the desired aiming light intersects the ICRN plane and the $x$-axis ,the desired aiming light beam and the ICRN original normal vector are in the same plane, namely horizontal plane. The plane consisting of the $y$-axis and the ICRN original normal vector is named as vertical plane.\par
From the geometric relationship, we can derive the light beam offset in the ICRN plane $R_k=\frac{tan\theta_kw_k}{cos\alpha}$, where $w_k$ is the path length from the transmitter to the ICRN and $\alpha$ is the incidence angle of the beam. Since $\theta_k$ is small, $R_k$ can be approximated as $\frac{\theta_kw_k}{cos\alpha}$. We decompose $R_k$ into $R_{k_x}, R_{k_y}$ along $\theta_k$ $x$ and $y$ axes in the ICRN plane, where $R_{k_x}^2+R_{k_y}^2=R_k^2$. Then $\theta_k$ can be decomposed into horizontal component $\theta _{x_k}$ and vertical component $\theta _{y_k}$ based on $R_{k_x}, R_{k_y}$, where $\theta_{x_k}=\frac{R_{k_x}cos\alpha}{w_k}, \theta_{y_k}=\frac{R_{k_y}cos\alpha}{w_k}$ . Both $\theta _{x_k}$, $\theta _{y_k}$ are subject to the standard normal distribution with probability density of \cite{borah2009pointing}\cite{arnon1997performance}\cite{farid2007outage}
\begin{equation} \label {3-1}
\begin{split}
f\left ( \theta _{x_{k}} \right )=\frac{1}{\sqrt{2\pi} \sigma _{\theta _{x_k}}}e^{-\frac{\theta _{x_k}^{2}}{2\sigma _{\theta _{x_k}}^{2}}}\\f\left ( \theta _{y_{k}} \right )=\frac{1}{\sqrt{2\pi} \sigma _{\theta _{y_k}}}e^{ -\frac{\theta _{y_k}^{2}}{2\sigma _{\theta _{y_k}}^{2}} }
\end{split}
\end{equation}
where $\sigma _{\theta _{x_k}}$ and $\sigma _{\theta _{y_k}}$ are the standard deviation of $\theta _{x_k}$ and $\theta _{y_k}$ respectively.\par
We use the deflection of the normal vector of the ICRN to describe the jitter of the ICRN plane. The direction of ICRN normal vector deflection can be decomposed into which in the horizontal plane and in the vertical plane, where the deflection angles are $\beta _{x_k}$, $\beta _{y_k}$ respectively. Based on the physical model of mirror jitter\cite{laughlin1995mirror}\cite{mcever2004adaptive}, we assume that both $\beta _{x_k}$, $\beta _{y_k}$ are subject to the standard normal distribution with probability density of
\begin{equation} \label{3-2}
\begin{split}
f\left ( \beta _{x_{k}} \right )=\frac{1}{\sqrt{2\pi} \sigma _{\beta _{x_k}}}e^{-\frac{\beta _{x_k}^{2}}{2\sigma _{\beta _{x_k}}^{2}}}\\f\left ( \beta _{y_{k}} \right )=\frac{1}{\sqrt{2\pi} \sigma _{\beta _{y_k}}}e^{ -\frac{\beta _{y_k}^{2}}{2\sigma _{\beta _{y_k}}^{2}}}
\end{split}
\end{equation}
where $\sigma _{\beta _{x_k}}$ and $\sigma _{\beta _{y_k}}$ are the standard deviation of $\beta _{x_k}$ and $\beta _{y_k}$ respectively.\par
By symmetry we can assume that
\begin{equation}
\begin{split} \label{3-4}
\sigma _{\theta _{x_k}}=\sigma _{\theta _{y_k}}=\sigma _{\theta _{k}},\\
\sigma _{\beta _{x_k}}=\sigma _{\beta _{y_k}}=\sigma _{\beta _{k}}.
\end{split}
\end{equation}
Below we will derive the relationship among $\theta_k, \beta_k$ and the superimposed pointing error angle $\theta_k^{(s)}$ in the horizontal and vertical plane respectively, where the superimposed pointing error angle $\theta_k^{(s)}$ is the angle formed by the receiver center, the ICRN reflection point and the actual incident point of the receiver. $\theta_k^{(s)}$ is the angle corresponding to the light beam offset in the receiver plane $r_k$ and can be decomposed into horizontal component $\theta _{x_k}^{(s)}$ and vertical component $\theta _{y_k}^{(s)}$ in the horizontal and vertical planes respectively.\par
\begin{figure}[htbp]
\centering
\includegraphics[width=1\textwidth]{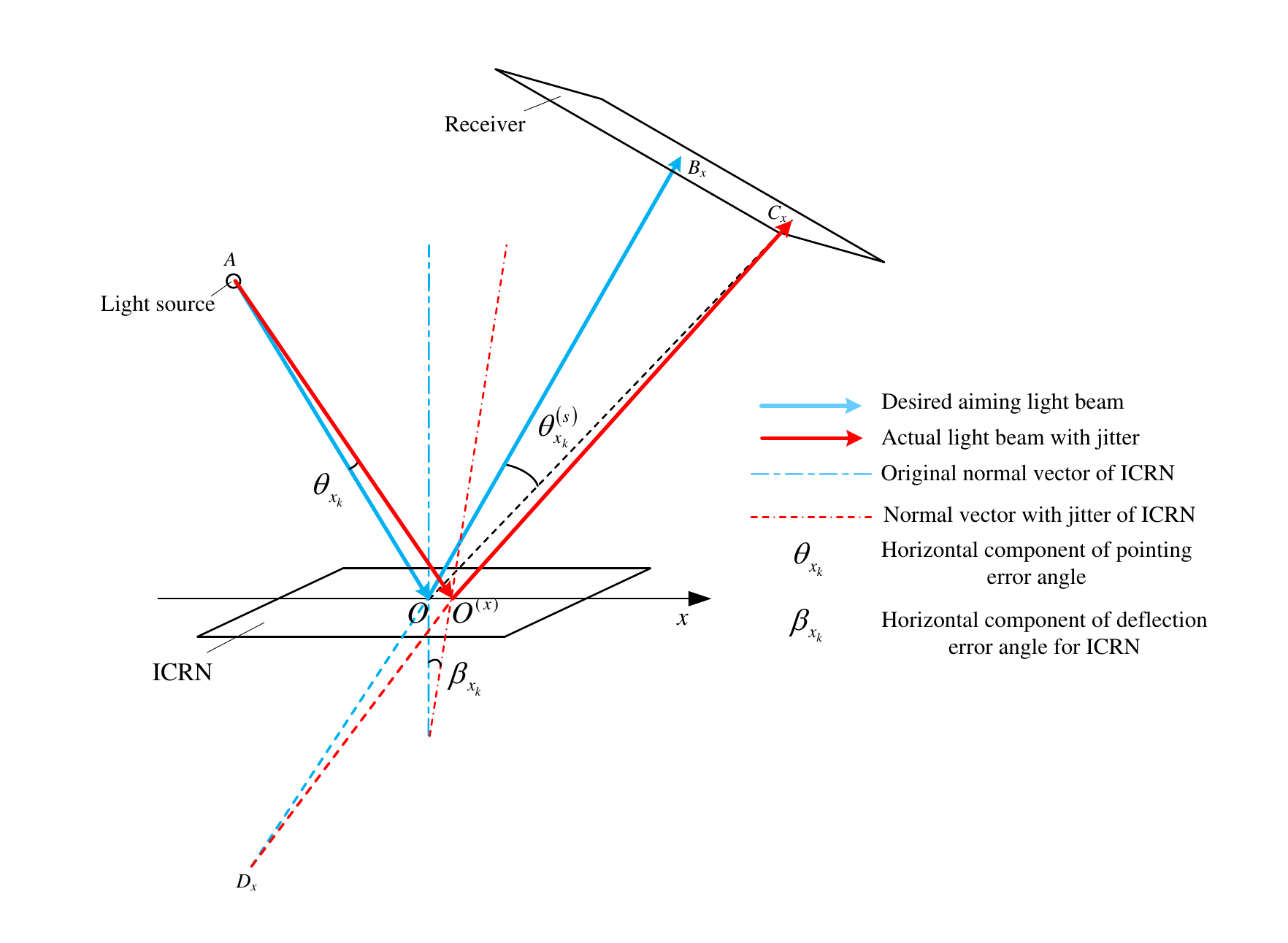}
\caption{Diagram of the optical intelligent channel system with single branch in the horizontal plane.}
\label {fig.3}
\end{figure}
\begin{figure}[htbp]
\centering
\includegraphics[width=1\textwidth]{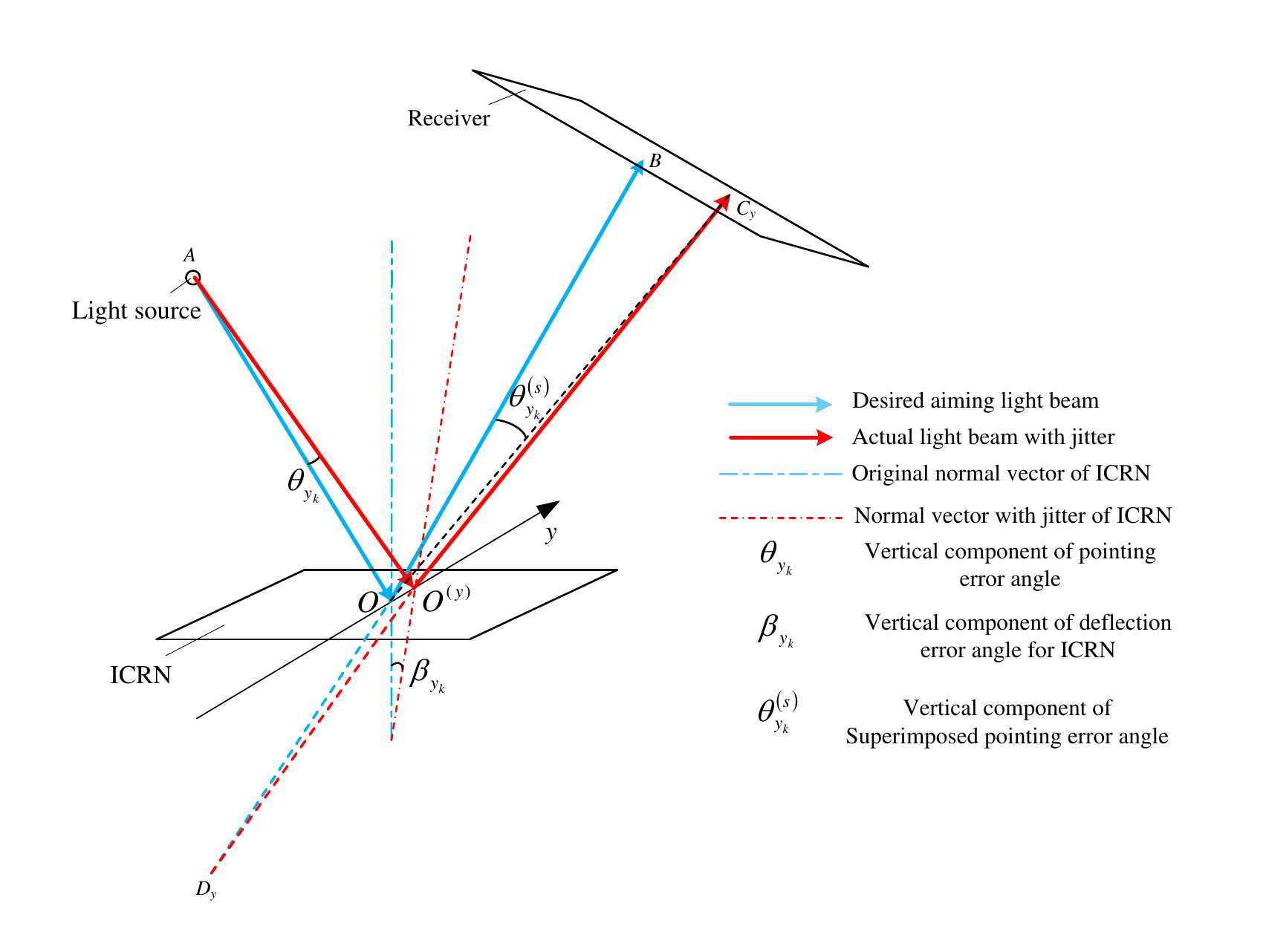}
\caption{Diagram of the optical intelligent channel system with single branch in the vertical plane.}
\label {fig.4}
\end{figure}

Fig. \ref{fig.3} shows the diagram of the optical intelligent channel system with single branch in the horizontal plane. We can derive the relationship among $\theta_{x_k}^{(s)}$ and $\theta_{x_k}, \beta_{x_k}$ according to Appendix \ref{appA} as
\begin{equation} \label{3-3-1}
\begin{aligned}
\theta_{x_k}^{(s)}\approx\left(1+\frac{w_k}{l_k}\right)\theta_{x_k}+2\beta_{x_k}.
\end{aligned}
\end{equation}
Fig. \ref{fig.4} shows the diagram of the optical intelligent channel system with single branch in the horizontal plane. The relationship among $\theta_{y_k}^{(s)}$ and $\theta_{y_k}, \beta_{y_k}$ can be derived according to Appendix \ref{appB} as
\begin{equation} \label{3-3-2}
\begin{aligned}
\theta_{y_k}^{(s)}\approx\left(1+\frac{w_k}{l_k}\right)\theta_{y_k}+2\beta_{y_k}.
\end{aligned}
\end{equation}
The superimposed pointing error angle $\theta^{(s)}_k$ is the root square sum of the horizontal and vertical angles and can be obtained as
\begin{equation} \label{3-5}
\begin{split}
\theta^{(s)}_k=\sqrt{\theta^{(s)2}_{x_k}+\theta^{(s)2}_{y_k}}.
\end{split}
\end{equation}
Since $\theta_{x_k}^{(s)}, \theta_{y_k}^{(s)}$ are independent and identically distributed, $\theta^{(s)}_k$ is subjected to the Rayleigh distribution with probability density of
\begin{equation} \label{3-6}
\begin{split}
f(\theta^{(s)}_k)=\frac{\theta^{(s)}_k}{\left(1+\frac{w_k}{l_k}\right)^2\sigma _{\theta _{k}}^{2}+  4\sigma _{\beta _{k}}^{2} }e^{-\frac{\theta^{(s)2}_k}{2\left(1+\frac{w_k}{l_k}\right)^2\sigma _{\theta _{k}}^{2}+  8\sigma _{\beta _{k}}^{2}}}.
\end{split}
\end{equation}
The cumulative distribution function(CDF) of $\theta^{(s)}_k$ is
\begin{equation} \label{3-7}
\begin{split}
F_{\theta^{(s)}_k}(x)=P(\theta^{(s)}_k\leq x)=1-exp\left(\frac{-x^2}{2\left(1+\frac{w_k}{l_k}\right)^2\sigma _{\theta _{k}}^{2}+  8\sigma _{\beta _{k}}^{2}}\right).
\end{split}
\end{equation}
\subsection{Probability of Obstruction}
In this section, we discuss the impact of obstacles on communication performance. We use a random variable $h_o$ to describe the channel fading caused by obstacles. For optical wireless communication, if the channel is blocked by an obstacle, $h_o=0$, the receiver can not receive any power through channel. If the channel is not blocked, $h_o=1$, the communication is not influenced by the obstacle.\par
For a free-space optical channel, we assume that the longer lasers are transmitted, the higher the probability of obstruction appears in the path. Suppose that in an optical channel of one unit length, the probability of obstruction appearing is $p_o$. Therefore, for an optical channel of $N$ unit length, the probability of obstruction appearing is $1-(1-p_o)^N$. Generalizing the observation to continuous channels, we can use $P_o=1-x^L$ to describe the probability of obstruction in the channel, where $L$ is the channel length, $x$ is a constant and $0<x<1$. In this paper, we assume $x=e^{-\eta}, \eta>0$. According to the relationship, $\eta$ is positively related to $P_o$. The PDF of $h_o$ can be presented as
\begin{equation} \label{4-1}
\begin{split}
f_{h_o}(h_o)=(1-e^{-\eta L})\delta(h_o)+e^{-\eta L}\delta(h_o-1)
\end{split}
\end{equation}
where $\delta(\cdot)$ is a unit-impulse function.
\subsection{Channel fading}
Since $\theta^{(s)}_k$ is the angle corresponding to the light beam offset in the receiver plane $r_k$, the instantaneous displacement from the receiver center to receiving light spot $r_k$ can be presented as
\begin{equation}\label{5-1}
\begin{split}
r_k=tan\theta^{(s)}_kl_k\approx\theta^{(s)}_kl_k.
\end{split}
\end{equation}
In this system, the Gaussian beam propagates through distance $(w_k+l_k)$ from the transmitter to the receiver with aperture radius $a$. The channel fading caused by pointing error can be approximated as \cite{farid2007outage}
\begin{equation} \label{5-2}
\begin{split}
h_{p_k} \approx A_0exp(-\frac{2r_k^2}{w_{zeq}^2})
\end{split}
\end{equation}
where $A_0$ is the fraction of the collected power at $r_k=0$, and $w_{zeq}$ is the equivalent beam width. We have $A_0=[erf(u)]^2$ and $w_{zeq}^2=w_z^2\frac{\sqrt{\pi}erf(u)}{2uexp(-u^2)}$, where $u=\sqrt{\frac{\pi}{2}}\frac{a}{w_z}$ is the ratio between aperture radius and beam width, and $erf(x)=\frac{2}{\sqrt{\pi}}\int_0^xe^{-t^2}dt$ is the error function. The beam width $w_z$ can be approximated by $w_z=\phi(l_k+w_k)$, where $\phi$ is the divergence angle of the beam, which describes the increase of the beam radius with the increase of the propagation distance from the transmitter. The approximation in \eqref{5-2} is accurate when $\frac{w_z}{a}>6$\cite{farid2007outage}. From \eqref{3-7} and \eqref{5-2}, we can obtain the PDF of $h_{p_k}$ as
\begin{equation} \label{5-3}
\begin{aligned}
f_{h_{p_k}}(h_{p_k})=\frac{m_k}{A_0}\left(\frac{h_{p_k}}{A_0}\right)^{m_k-1},\quad 0<h_{p_k}<A_0
\end{aligned}
\end{equation}
where
\begin{equation} \label{5-4}
\begin{aligned}
m_k=\frac{w_{zeq}^2}{4\sigma _{\theta _{k}}^{2}\left(l_k+w_k\right)^2+16\sigma _{\beta _{k}}^{2}l_k^2}.
\end{aligned}
\end{equation}
Considering the probability of an obstacle, we can obtain the channel power fading of the $k$th channel as
\begin{equation} \label{5-5}
\begin{aligned}
h_k=h_{p_k}h_{o_k}=A_0exp(-\frac{2\theta^{(s)2}_kl_k^2}{w_{zeq}^2})h_{o_k}.
\end{aligned}
\end{equation}
The CDF of the channel power fading can be presented as
\begin{equation} \label{5-6}
\begin{aligned}
F_{h_k}(x)&=\iint_{h_{p_k}h_{o_k}\leq x}f_{h_{p_k}}(h_{p_k})f_{h_{o_k}}(h_{o_k})dh_{p_k}dh_{o_k}\\&=\int_0^{A_0}\int_0^{\frac{x}{h_p}}\frac{m_k(1-e^{-\eta_k(l_k+w_k)})}{A_0}\left(\frac{h_{p_k}}{A_0}\right)^{m_k-1}\delta(h_o)+\frac{m_ke^{-\eta_k(l_k+w_k)}}{A_0}\left(\frac{h_{p_k}}{A_0}\right)^{m_k-1}\delta(h_o-1)dh_odh_k
\\&=\left\{\begin{matrix}1-n_k+n_k\left(\frac{x}{A_0}\right)^{m_k},\ &0<x\leq A_0\\1,\ &x>A_0\end{matrix}\right.
\end{aligned}
\end{equation}
where $1-n_k=1-e^{-\eta_k(l_k+w_k)}$ is probability of an obstacle in the $k$th channel. Taking derivative of \eqref{5-6}, we can obtain the PDF of $h_k$ as \cite{phillips2003signals}
\begin{equation} \label{5-7}
\begin{aligned}
f_{h_k}(h_k)=(1-n_k)\delta(h_k)+n_k\frac{m_k}{A_0}\left(\frac{h_{k}}{A_0}\right)^{m_k-1}, \quad 0<h_{k}<A_0.
\end{aligned}
\end{equation}
\section{Error Rate and Outage Probability Performance}\label{analysis}
\subsection{Summary of Asymptotic Analysis Techniques}
Our derivation process requires results from \cite{wang2003simple}, which we should recall in this section. We can decompose the SNR of the system $\gamma$ into $\gamma=\overline{\gamma}\mu$, where $\overline{\gamma}$ represents the average SNR and $\mu$ is a random variable. Suppose that the PDF of $\mu$ is
\begin{equation} \label{6-1}
\begin{split}
f_{\mu}(\mu)=g_c\mu^{t}+o(\mu^t)
\end{split}
\end{equation}
where $g_c\mu^{t}$ is the first non-zero term of $f_{\mu}(\mu)$ Taylor series expansion at zero, $o(\mu^t)$ is the higher-order term. The PDF of $\gamma$ can be presented as
\begin{equation} \label{6-2}
\begin{aligned}
f_{\gamma}(\gamma)=\frac{g_c\gamma^t}{\overline{\gamma}^{t+1}}+o(\gamma^t).
\end{aligned}
\end{equation}
The outage probability ,which is defined in \cite{ko2000outage}, can be presented as
\begin{equation} \label{6-3}
\begin{aligned}
P_{out}(\gamma_{th})&=\int_{0 }^{\gamma_{th} }f_{\gamma}(\gamma)d\gamma\\&=\frac{g_c}{t+1}\left(\frac{\gamma_{th}}{\overline{\gamma}}\right)^{t+1}+o\left(\frac{1}{\overline{\gamma}^{t+1}}\right).
\end{aligned}
\end{equation}
The average BER of the coherent modulation scheme with conditional error rate $P_e(\mu)=\rho Q(\sqrt{\overline{\gamma}\zeta\mu})$, where $Q(\cdot)$ is the Gaussian function, $\rho$ and $\zeta$ are constants associated with the underlying modulation format, is derived by \cite{wang2003simple} as
\begin{equation} \label{6-4}
\begin{aligned}
P_{e}&=\int_{0}^{\infty}\rho Q(\sqrt{\overline{\gamma}\zeta\mu})f_{\mu}(\mu)d\mu\\&=\frac{2^tg_c\rho\Gamma\left(t+\frac{3}{2}\right)}{\sqrt{\pi}(t+1)(\zeta\overline{\gamma})^{t+1}}+o\left(\frac{1}{\overline{\gamma}^{t+1}}\right)
\end{aligned}
\end{equation}
where $\Gamma(\cdot)$ is the gamma function. When it is difficult to obtain the PDF of SNR, we can use the moment generating function (MGF) to obtain $g_c$ and $t$, which can be presented as
\begin{equation} \label{6-5}
\begin{aligned}
M_{\gamma}(v)&=E\left[e^{-v\gamma}\right]\\&=\int_{0}^{\infty}e^{-v\gamma}f_{\gamma}(\gamma)d\gamma\\&=\frac{g_c\Gamma(t+1)}{\overline{\gamma}^{t+1}v^{t+1}}+o\left(\frac{1}{v^{t+1}}\right)
\end{aligned}
\end{equation}
where $E\left[\cdot\right]$ represents expectation. After obtaining $g_c$ and $t$ from \eqref{6-1} or \eqref{6-5}, we can obtain the asymptotic outage probability and BER according to \eqref{6-3} and \eqref{6-4}.
\subsection{Asymptotic Performance Analysis of Single-branch System}
According to \eqref{2-1}, we can assume $\alpha_k=1$ and $E\left[s_k^2\right]=2P_t^2$ for single-branch case, and the instantaneous SNR in $k$th channel $\gamma_k$ can be defined as\cite{farid2007outage}\cite{yang2014free}
\begin{equation} \label{7-1}
\begin{split}
\gamma_k=\frac{2P_t^2\alpha_k^2h_k^2}{\sigma_n^2}=\frac{2P_t^2h_k^2}{\sigma_n^2}
\end{split}
\end{equation}
Substituting \eqref{7-1} into \eqref{5-6}, we can obtain the CDF of $\gamma_k$ as
\begin{equation} \label{7-2}
\begin{aligned}
F_{\gamma_k}(x)&=F_{h_k}(\sqrt{\frac{\sigma_n^2x}{2P_t^2}})\\&=\left\{\begin{matrix}1-n_k+n_k\left(\frac{\sigma_n^2x}{2P_t^2A_0^2}\right)^{\frac{m_k}{2}},\ &0<x\leq \frac{2P_t^2A_0^2}{\sigma_n^2}\\ 1,\ &x>\frac{2P_t^2A_0^2}{\sigma_n^2}\end{matrix}\right..
\end{aligned}
\end{equation}
Then the PDF of $\gamma_k$ can be presented as
\begin{equation} \label{7-3}
\begin{aligned}
f_{\gamma_k}(\gamma_k)=(1-n_k)\delta(\gamma_k)+\frac{m_kn_k}{2}\left(\frac{\sigma_n^2}{2P_t^2A_0^2}\right)^{\frac{m_k}{2}}\gamma_k^{\frac{m_k}{2}-1},\quad 0<\gamma_k<\frac{2P_t^2A_0^2}{\sigma_n^2}.
\end{aligned}
\end{equation}
Let $\gamma_k=\overline{\gamma_k}\mu_k$, where $\overline{\gamma_k}=\frac{2P_t^2}{\sigma_n^2}$ represents the average SNR of the kth channel, $\mu_k=h_k^2$ is a channel-dependent random variable (RV). Then the PDF of $\mu_k$ is
\begin{equation} \label{7-4}
\begin{aligned}
f_{\mu_k}(\mu_k)&=(1-n_k)\delta(\mu_k)+\frac{m_kn_k}{2A_0^2}\left(\frac{\mu_k}{A_0^2}\right)^{\frac{m_k}{2}-1}\\&=(1-n_k)\delta(\mu_k)+g_{c_k}\mu_k^{t_k}, \qquad 0<\mu_k<A_0^2
\end{aligned}
\end{equation}
where
\begin{equation} \label{7-5}
\begin{aligned}
g_{c_k}=\frac{m_kn_k}{2A_0^{m_k}},\quad t_k=\frac{m_k}{2}-1.
\end{aligned}
\end{equation}
For IM/DD with OOK modulation, the conditional error rate is $P_e(\mu_k)=Q(\sqrt{\frac{1}{2}\overline{\gamma_k}\mu_k})$. The average BER of the $k$th channel can be obtained as
\begin{equation} \label{7-6}
\begin{aligned}
P_{e_k}&=\int_0^{\infty} Q(\sqrt{\frac{1}{2}\overline{\gamma_k}\mu_k})f_{\mu_k}(\mu_k)d\mu_k\\&=\int_0^\infty (1-n_k)Q(\sqrt{\frac{1}{2}\overline{\gamma_k}\mu_k})\delta(\mu_k)d\mu_k+\int_0^\infty \frac{m_kn_k}{2A_0^{m_k}}Q(\sqrt{\frac{1}{2}\overline{\gamma_k}\mu_k})\mu_k^{\frac{m_k}{2}-1}d\mu_k\\&=\frac{1-n_k}{2}+\frac{n_k\left(\frac{2\sigma_n^2}{P_t^2A_0^2}\right)^{\frac{m_k}{2}}\gamma\left(\frac{m_k+1}{2},\overline{\gamma_k}A_0^2\right)}{2\sqrt{\pi}}.
\end{aligned}
\end{equation}
where $\gamma(v,z)$ is an incomplete gamma function and $\gamma(v,z)=\int_0^zu^{v-1}e^{-u}du$. When $\gamma_k\to \infty$, we can obtain the asymptotic average BER of the $k$th channel as
\begin{equation} \label{7-6-2}
\begin{aligned}
P_{e_k}^{\infty}=\frac{1-n_k}{2}+\frac{n_k\left(\frac{2\sigma_n^2}{P_t^2A_0^2}\right)^{\frac{m_k}{2}}\Gamma\left(\frac{m_k+1}{2}\right)}{2\sqrt{\pi}}.
\end{aligned}
\end{equation}
We can observe from \eqref{7-6} and \eqref{7-6-2} that there exists an error rate floor, which is equal to $\frac{1-n_k}{2}$.\par
The outage probability of $k$th channel can be obtained as
\begin{equation} \label{7-7}
\begin{aligned}
P_{out_k}(\gamma_{th})&=\int_{0 }^{\gamma_{th} }f_{\gamma_k}(\gamma_k)d\gamma_k\\&=\int_{0}^{\gamma_{th}}(1-n_k)\delta(\gamma_k)d\gamma_k+\int_{0}^{\gamma_{th}}\frac{m_kn_k}{2}\left(\frac{\sigma_n^2}{2P_t^2A_0^2}\right)^{\frac{m_k}{2}}\gamma_k^{\frac{m_k}{2}-1}d\gamma_k
\\&=\left\{\begin{matrix}1-n_k+n_k\left(\frac{\sigma_n^2\gamma_{th}}{2P_t^2A_0^2}\right)^{\frac{m_k}{2}},\ &0<x\leq \frac{2P_t^2A_0^2}{\sigma_n^2}\\1,\ &x>\frac{2P_t^2A_0^2}{\sigma_n^2}\end{matrix}\right.
\end{aligned}
\end{equation}
where $\gamma_{th}$ is the outage threshold. It can be seen from \eqref{7-7} that there exists an outage probability floor, which is equal to $1-n_k$\par
The asymptotic MGF of $\gamma_k$ can be derived from \eqref{6-5} as
\begin{equation} \label{7-8}
\begin{aligned}
M_{\gamma_k}(v)=E\left[e^{-v\gamma_k}\right]&=\int_{0}^{\infty}e^{-v\gamma_k}f_{\gamma_k}(\gamma_k)d\gamma_k\\&=\int_0^\infty (1-n_k)e^{-v\gamma_k}\delta(\gamma_k)d\gamma_k+\int_0^\infty \frac{m_kn_k}{2}\left(\frac{\sigma_n^2}{2P_t^2A_0^2}\right)^{\frac{m_k}{2}}\gamma_k^{\frac{m_k}{2}-1}e^{-v\gamma_k}d\gamma_k\\&=1-n_k+\frac{m_kn_k}{2}\left(\frac{\sigma_n^2}{2P_t^2A_0^2v}\right)^{\frac{m_k}{2}}\Gamma(\frac{m_k}{2}).
\end{aligned}
\end{equation}
\subsection{Asymptotic Performance Analysis of Multi-branch System}
Considering the M-branch case, the total transmitted power is allocated according to the power allocation coefficient $\alpha_k$. At the receiving end, we utilize maximum ratio combining (MRC) and obtain the SNR of the intelligent channel system as
\begin{equation} \label{7-9}
\begin{aligned}
\gamma=\sum_{k=0}^{M-1}\alpha_k^2\gamma_k.
\end{aligned}
\end{equation}
Suppose that $\gamma_k$ for different channels are independent in this system, then the asymptotic MGF of $\gamma$ is
\begin{equation} \label{7-10}
\begin{aligned}
M_{\gamma}(v)&=E\left[e^{-v\gamma}\right]\\&=E\left[e^{-v\sum_{k=0}^{M-1}\alpha_k^2\gamma_k}\right]\\&=E\left[e^{-v\alpha_0^2\gamma_0}\right]E\left[e^{-v\alpha_1^2\gamma_1}\right]\cdots E\left[e^{-v\alpha_{M-1}^2\gamma_{M-1}}\right]\\&=\prod_{k=0}^{M-1}M_{\gamma_k}(\alpha _{k}^2v)\\&=\prod_{k=0}^{M-1}\left[1-n_k+\frac{m_kn_k}{2}\left(\frac{\sigma_n^2}{2P_t^2A_0^2\alpha_k^2v}\right)^{\frac{m_k}{2}}\Gamma(\frac{m_k}{2})\right].
\end{aligned}
\end{equation}
We need to expand $M_{\gamma}(v)$ and integrate each term to derive the asymptotic PDF of $\gamma$, thus for this higher-order polynomial we need to discard some terms to simplify the expression. When the system works at high SNR, the higher-order terms of $(\frac{2P_t^2A_0^2}{\sigma_n^2})^{\frac{m_k}{2}}$ can be discarded for formula simplification. When the system works at lower SNR and the probability of obstruction is relatively small, the higher-order terms of $1-n_k$ can be discarded for formula simplification. Therefore, in order to make the expression satisfy various situations, we keep both the zeroth and the first-order terms of $1-n_k$ and $(\frac{2P_t^2A_0^2}{\sigma_n^2})^{\frac{m_k}{2}}$. Therefore, the asymptotic $M_{\gamma}(v)$ can be approximated as
\begin{equation} \label{7-11}
\begin{aligned}
M_{\gamma}(v)\approx&\prod_{k=0}^{M-1}(1-n_k)+\sum_{k=0}^{M-1}\frac{m_kn_k}{2}\left(\frac{\sigma_n^2}{2P_t^2A_0^2\alpha_k^2v}\right)^{\frac{m_k}{2}}\Gamma(\frac{m_k}{2})\frac{\prod_{i=0}^{M-1}(1-n_i)}{1-n_k}
\\+&\sum_{k=0}^{M-1}(1-n_k)\frac{\prod_{i=0}^{M-1}\frac{m_in_i}{2}\left(\frac{\sigma_n^2}{2P_t^2A_0^2\alpha_i^2v}\right)^{\frac{m_i}{2}}\Gamma(\frac{m_i}{2})}
{\frac{m_kn_k}{2}\left(\frac{\sigma_n^2}{2P_t^2A_0^2\alpha_k^2v}\right)^{\frac{m_k}{2}}\Gamma(\frac{m_k}{2})}
+\prod_{k=0}^{M-1}\frac{m_kn_k}{2}\left(\frac{\sigma_n^2}{2P_t^2A_0^2\alpha_k^2v}\right)^{\frac{m_k}{2}}\Gamma(\frac{m_k}{2}).
\end{aligned}
\end{equation}
Then we can obtain the asymptotic PDF of $\gamma$ as
\begin{equation} \label{7-12}
\begin{aligned}
f_{\gamma}(\gamma)&=\int_{-\infty}^{\infty}M_{\gamma}(v)e^{v\gamma}dv
\\&=\int_{-\infty}^{\infty}\prod_{k=0}^{M-1}e^{v\gamma}(1-n_k)dv+\sum_{k=0}^{M-1}\int_{-\infty}^{\infty}e^{v\gamma}\frac{m_kn_k}{2}\left(\frac{\sigma_n^2}{2P_t^2A_0^2\alpha_k^2v}\right)^{\frac{m_k}{2}}\Gamma(\frac{m_k}{2})\frac{\prod_{i=0}^{M-1}(1-n_i)}{1-n_k}dv
\\&+\sum_{k=0}^{M-1}\int_{-\infty}^{\infty}(1-n_k)\frac{\prod_{i=0}^{M-1}\frac{m_in_i}{2}\left(\frac{\sigma_n^2}{2P_t^2A_0^2\alpha_i^2v}\right)^{\frac{m_i}{2}}\Gamma(\frac{m_i}{2})}
{\frac{m_kn_k}{2}\left(\frac{\sigma_n^2}{2P_t^2A_0^2\alpha_k^2v}\right)^{\frac{m_k}{2}}\Gamma(\frac{m_k}{2})}dv
+\int_{-\infty}^{\infty}\prod_{k=0}^{M-1}\frac{m_kn_k}{2}\left(\frac{\sigma_n^2}{2P_t^2A_0^2\alpha_k^2v}\right)^{\frac{m_k}{2}}\Gamma(\frac{m_k}{2})dv
\\&=\prod_{k=0}^{M-1}(1-n_k)\delta(\gamma)+\sum_{k=0}^{M-1}\frac{m_kn_k}{2}\gamma^{\frac{m_k}{2}-1}\left(\frac{\sigma_n^2}{2P_t^2A_0^2\alpha_k^2}\right)^{\frac{m_k}{2}}\frac{\prod_{i=0}^{M-1}(1-n_i)}{1-n_k}
\\&+\sum_{k=0}^{M-1}(1-n_k)\gamma^{\frac{m-m_k}{2}-1}\frac{\prod_{i=0}^{M-1}\frac{m_in_i}{2}\left(\frac{\sigma_n^2}{2P_t^2A_0^2\alpha_i^2}\right)^{\frac{m_i}{2}}\Gamma(\frac{m_i}{2})}
{\frac{m_kn_k}{2}\left(\frac{\sigma_n^2}{2P_t^2A_0^2\alpha_k^2}\right)^{\frac{m_k}{2}}\Gamma(\frac{m_k}{2})\Gamma(\frac{m-m_k}{2})}
+\frac{\gamma^{\frac{m}{2}-1}}{\Gamma(\frac{m}{2})}\prod_{k=0}^{M-1}\frac{m_kn_k}{2}\left(\frac{\sigma_n^2}{2P_t^2A_0^2\alpha_k^2}\right)^{\frac{m_k}{2}}\Gamma(\frac{m_k}{2})
\end{aligned}
\end{equation}
where $m=\sum_{k=0}^{M-1}m_k$. For IM/DD with OOK modulation, as the system's conditional error rate is $P_e(\gamma)=Q(\sqrt{\frac{1}{2}\gamma})$, the asymptotic average BER can be written as
\begin{equation} \label{7-13}
\begin{aligned}
P_{e}&=\int_{-\infty}^{\infty}Q(\sqrt{\frac{1}{2}\gamma})f_{\gamma}(\gamma)d\gamma\\
&=\int_{-\infty}^{\infty}Q(\sqrt{\frac{1}{2}\gamma})\prod_{k=0}^{M-1}(1-n_k)\delta(\gamma)d\gamma+\sum_{k=0}^{M-1}\int_{-\infty}^{\infty}Q(\sqrt{\frac{1}{2}\gamma})\frac{m_kn_k}{2}\gamma^{\frac{m_k}{2}-1}\left(\frac{\sigma_n^2}{2P_t^2A_0^2\alpha_k^2}\right)^{\frac{m_k}{2}}\frac{\prod_{i=0}^{M-1}(1-n_i)}{1-n_k}d\gamma
\\&+\sum_{k=0}^{M-1}\int_{-\infty}^{\infty}(1-n_k)Q(\sqrt{\frac{1}{2}\gamma})\gamma^{\frac{m-m_k}{2}-1}\frac{\prod_{i=0}^{M-1}\frac{m_in_i}{2}\left(\frac{\sigma_n^2}{2P_t^2A_0^2\alpha_i^2}\right)^{\frac{m_i}{2}}\Gamma(\frac{m_i}{2})}
{\frac{m_kn_k}{2}\left(\frac{\sigma_n^2}{2P_t^2A_0^2\alpha_k^2}\right)^{\frac{m_k}{2}}\Gamma(\frac{m_k}{2})\Gamma(\frac{m-m_k}{2})}d\gamma
\\&+\int_{-\infty}^{\infty}Q(\sqrt{\frac{1}{2}\gamma})\frac{\gamma^{\frac{m}{2}-1}}{\Gamma(\frac{m}{2})}\prod_{k=0}^{M-1}\frac{m_kn_k}{2}\left(\frac{\sigma_n^2}{2P_t^2A_0^2\alpha_k^2}\right)^{\frac{m_k}{2}}\Gamma(\frac{m_k}{2})d\gamma
\\&=\frac{1}{2}\prod_{k=0}^{M-1}(1-n_k)+\sum_{k=0}^{M-1}m_kn_k2^{m_k-1}\left(\frac{\sigma_n^2}{2P_t^2A_0^2\alpha_k^2}\right)^{\frac{m_k}{2}}\Gamma(\frac{m_k+1}{2})\frac{\prod_{i=0}^{M-1}(1-n_i)}{\sqrt{\pi}(1-n_k)m_k}
\\&+\sum_{k=0}^{M-1}(1-n_k)\frac{2^{m-m_k}\Gamma(\frac{m-m_k+1}{2})\prod_{i=0}^{M-1}\frac{m_in_i}{2}\left(\frac{\sigma_n^2}{2P_t^2A_0^2\alpha_i^2}\right)^{\frac{m_i}{2}}\Gamma(\frac{m_i}{2})}
{\frac{\sqrt{\pi}m_kn_k(m-m_k)}{2}\left(\frac{\sigma_n^2}{2P_t^2A_0^2\alpha_k^2}\right)^{\frac{m_k}{2}}\Gamma(\frac{m_k}{2})\Gamma(\frac{m-m_k}{2})}
+\frac{2^{m}\Gamma(\frac{m+1}{2})}{\sqrt{\pi}m\Gamma(\frac{m}{2})}\prod_{k=0}^{M-1}\frac{m_kn_k}{2}\left(\frac{\sigma_n^2}{2P_t^2A_0^2\alpha_k^2}\right)^{\frac{m_k}{2}}\Gamma(\frac{m_k}{2}).
\end{aligned}
\end{equation}
The asymptotic outage probability of the system is
\begin{equation} \label{7-14}
\begin{aligned}
P_{out}(\gamma_{th})&=\int_{0}^{\gamma_{th}}f_{\gamma}(\gamma)d\gamma\\&=\int_{0}^{\gamma_{th}}\prod_{k=0}^{M-1}(1-n_k)\delta(\gamma)d\gamma+\sum_{k=0}^{M-1}\int_{0}^{\gamma_{th}}\frac{m_kn_k}{2}\gamma^{\frac{m_k}{2}-1}\left(\frac{\sigma_n^2}{2P_t^2A_0^2\alpha_k^2}\right)^{\frac{m_k}{2}}\frac{\prod_{i=0}^{M-1}(1-n_i)}{1-n_k}d\gamma
\\&+\sum_{k=0}^{M-1}\int_{0}^{\gamma_{th}}(1-n_k)\gamma^{\frac{m-m_k}{2}-1}\frac{\prod_{i=0}^{M-1}\frac{m_in_i}{2}\left(\frac{\sigma_n^2}{2P_t^2A_0^2\alpha_i^2}\right)^{\frac{m_i}{2}}\Gamma(\frac{m_i}{2})}
{\frac{m_kn_k}{2}\left(\frac{\sigma_n^2}{2P_t^2A_0^2\alpha_k^2}\right)^{\frac{m_k}{2}}\Gamma(\frac{m_k}{2})\Gamma(\frac{m-m_k}{2})}d\gamma
\\&+\int_{0}^{\gamma_{th}}\frac{\gamma^{\frac{m}{2}-1}}{\Gamma(\frac{m}{2})}\prod_{k=0}^{M-1}\frac{m_kn_k}{2}\left(\frac{\sigma_n^2}{2P_t^2A_0^2\alpha_k^2}\right)^{\frac{m_k}{2}}\Gamma(\frac{m_k}{2})d\gamma
\\&=\prod_{k=0}^{M-1}(1-n_k)+\sum_{k=0}^{M-1}\left(\frac{\sigma_n^2\gamma_{th}}{2P_t^2A_0^2\alpha_k^2}\right)^{\frac{m_k}{2}}\frac{n_k\prod_{i=0}^{M-1}(1-n_i)}{1-n_k}
\\&+\sum_{k=0}^{M-1}(1-n_k)\gamma_{th}^{\frac{m-m_k}{2}}\frac{\prod_{i=0}^{M-1}\frac{m_in_i}{2}\left(\frac{\sigma_n^2}{2P_t^2A_0^2\alpha_i^2}\right)^{\frac{m_i}{2}}\Gamma(\frac{m_i}{2})}
{\frac{m_kn_k(m-m_k)}{4}\left(\frac{\sigma_n^2}{2P_t^2A_0^2\alpha_k^2}\right)^{\frac{m_k}{2}}\Gamma(\frac{m_k}{2})\Gamma(\frac{m-m_k}{2})}
\\&+\frac{2\gamma_{th}^{\frac{m}{2}}}{m\Gamma(\frac{m}{2})}\prod_{k=0}^{M-1}\frac{m_kn_k}{2}\left(\frac{\sigma_n^2}{2P_t^2A_0^2\alpha_k^2}\right)^{\frac{m_k}{2}}\Gamma(\frac{m_k}{2}).
\end{aligned}
\end{equation}
\section{General Discussion}\label{discussion}
\subsection{Performance gain by adding an intelligent channel}
In order to quantify the performance gain after adding a channel, we suppose that all intelligent channels in the system are exactly the same and analyze the performance gain of adding one channel when there are $N$ channels in the system. According to the assumption that the state of each channel in the system is exactly the same, the transmitting end distributes power evenly to each channel. The average BER of system with $N$ identical channels can be derived from \eqref{7-13} as
\begin{equation} \label{9-1}
\begin{aligned}
P_{e}^{(N)}&=\frac{1}{2}(1-n_k)^{N}+Nm_kn_k2^{m_k-1}\left(\frac{\sigma_n^2N^2}{2P_t^2A_0^2}\right)^{\frac{m_k}{2}}\Gamma(\frac{m_k+1}{2})\frac{(1-n_k)^{N-1}}{\sqrt{\pi}m_k}
\\&+N(1-n_k)\frac{2^{(N-1)m_k}\Gamma(\frac{(N-1)m_k+1}{2})(\frac{m_kn_k}{2})^{N-1}\left(\frac{\sigma_n^2N^2}{2P_t^2A_0^2}\right)^{\frac{(N-1)m_k}{2}}(\Gamma(\frac{m_k}{2}))^{N-1}}
{\sqrt{\pi}(N-1)m_k\Gamma(\frac{(N-1)m_k}{2})}
\\&+\frac{2^{Nm_k}\Gamma(\frac{Nm_k+1}{2})}{\sqrt{\pi}Nm_k\Gamma(\frac{Nm_k}{2})}(\frac{m_kn_k}{2})^{N}\left(\frac{\sigma_n^2N^2}{2P_t^2A_0^2}\right)^{\frac{Nm_k}{2}}(\Gamma(\frac{m_k}{2}))^{N}.
\end{aligned}
\end{equation}
The BER performance gain after adding an intelligent channel can be presented as
\begin{equation} \label{9-2}
\begin{aligned}
g^{(N)}=\frac{P_{e}^{(N)}}{P_{e}^{(N+1)}}.
\end{aligned}
\end{equation}
By analyzing the relationship between $g^{(N)}$ and $N$, we can obtain the asymptotic performance gain brought by increasing of the number of intelligent channels.
\subsubsection{Performance gain at infinite SNR}
From \eqref{9-1}, we can observe that when $P_t \to \infty$, $P_{e}^{(N)} \to \frac{1}{2}(1-n_k)^{N}$, thus $g^{(N)} \to \frac{1}{1-n_k}$. Therefore, at infinite SNR, each additional intelligent channel can effectively reduce the BER floor caused by the probability of obstruction and the performance gain at infinite SNR is unrelated to $N$.
\subsubsection{Performance gain with low probability of obstruction}
From \eqref{9-1}, we can observe that when $1-n_k \to 0$, $P_{e}^{(N)} \to \frac{2^{Nm_k}\Gamma(\frac{Nm_k+1}{2})}{\sqrt{\pi}Nm_k\Gamma(\frac{Nm_k}{2})}(\frac{m_kn_k}{2})^{N}\left(\frac{\sigma_n^2N^2}{2P_t^2A_0^2}\right)^{\frac{Nm_k}{2}}(\Gamma(\frac{m_k}{2}))^{N}$, thus
\begin{equation} \label{9-2-2}
\begin{aligned}
g^{(N)}\to \frac{\Gamma(\frac{(N+1)m_k}{2})\Gamma(\frac{Nm_k+1}{2})N^{Nm_k-1}}{\Gamma(\frac{Nm_k}{2})\Gamma(\frac{(N+1)m_k+1}{2})(N+1)^{(N+1)m_k-1}m_kn_k2^{m_k-1}(\frac{\sigma_n^2}{2P_t^2A_0^2})^{\frac{m_k}{2}}\Gamma(\frac{m_k}{2})}.
\end{aligned}
\end{equation}
With low probability of obstruction, as $N$ increases, the performance gain $g^{(N)}$ tends to decrease. \par For further investigation, the performance gain curve at specific SNR with different probability of obstruction will be presented in Section \ref{simulation}, which is based on the relationship between $g^{(N)}$ and $N$.
\subsection{Power allocation scheme at high SNR}
In the practical scenario, since the channel state information (CSI) of each intelligent channel is different, the power allocated by the transmitting end to each channel should also be different, which is reflected in the power allocation coefficient $\alpha_k$ at transmitter in \eqref{2-1}. We take the system's BER as the objective function and minimize the BER by adjusting the value of $\alpha_k$. Therefore, the following optimization equation can be obtained as
\begin{equation} \label{d-1}
\begin{aligned}
\left\{\begin{matrix}\min &\quad P_e
\\
s.t.& \quad \sum_{k=0}^{M-1}\alpha_k=1.
\end{matrix}\right.
\end{aligned}
\end{equation}
Here we discuss the power allocation scheme in the case of high SNR, so the higher-order terms of $(\frac{2P_t^2A_0^2}{\sigma_n^2})^{\frac{m_k}{2}}$ in \eqref{7-13} can be omitted to simplify the expression, where the average BER can be approximated as
\begin{equation} \label{d-2}
\begin{aligned}
P_{e}^{\infty}\approx\frac{1}{2}\prod_{k=0}^{M-1}(1-n_k)+\sum_{k=0}^{M-1}m_kn_k2^{m_k-1}\left(\frac{\sigma_n^2}{2P_t^2A_0^2\alpha_k^2}\right)^{\frac{m_k}{2}}\Gamma(\frac{m_k+1}{2})\frac{\prod_{i=0}^{M-1}(1-n_i)}{\sqrt{\pi}(1-n_k)m_k}.
\end{aligned}
\end{equation}
The solution to the optimization problem in \eqref{d-1} can be obtained according to Appendix \ref{PowerAllocationProof} as
\begin{equation} \label{d-3}
\begin{aligned}
\alpha_i=\frac{(b_im_i)^{\frac{1}{m_i+1}}}{\sum_{j=0}^{M-1}(b_jm_j)^{\frac{1}{m_j+1}}}, \quad\quad i=0,1,2,\cdots,M-1
\end{aligned}
\end{equation}
where
\begin{equation} \label{d-4}
\begin{aligned}
b_i=m_in_i2^{m_i-1}\left(\frac{\sigma_n^2}{2P_t^2A_0^2}\right)^{\frac{m_i}{2}}\Gamma(\frac{m_i+1}{2})\frac{\prod_{j=0}^{M-1}(1-n_j)}{\sqrt{\pi}(1-n_i)m_i}, \quad\quad i=0,1,2,\cdots,M-1.
\end{aligned}
\end{equation}
\section{Numerical Results}\label{simulation}
In this section, we utilize the analytical results to study the performance of the intelligent channel system and the simulation results are used to demonstrate the analytical results. Firstly we will verify the CDF expression of the pointing error displacement $r_k$ in an intelligent channel, which is the basis of the analysis of channel fading for intelligent channels.
\subsection{Simulation and Analysis of the Pointing Error in Intelligent Channel}
\begin{figure}[htbp]
\centering
\includegraphics[width=1\textwidth]{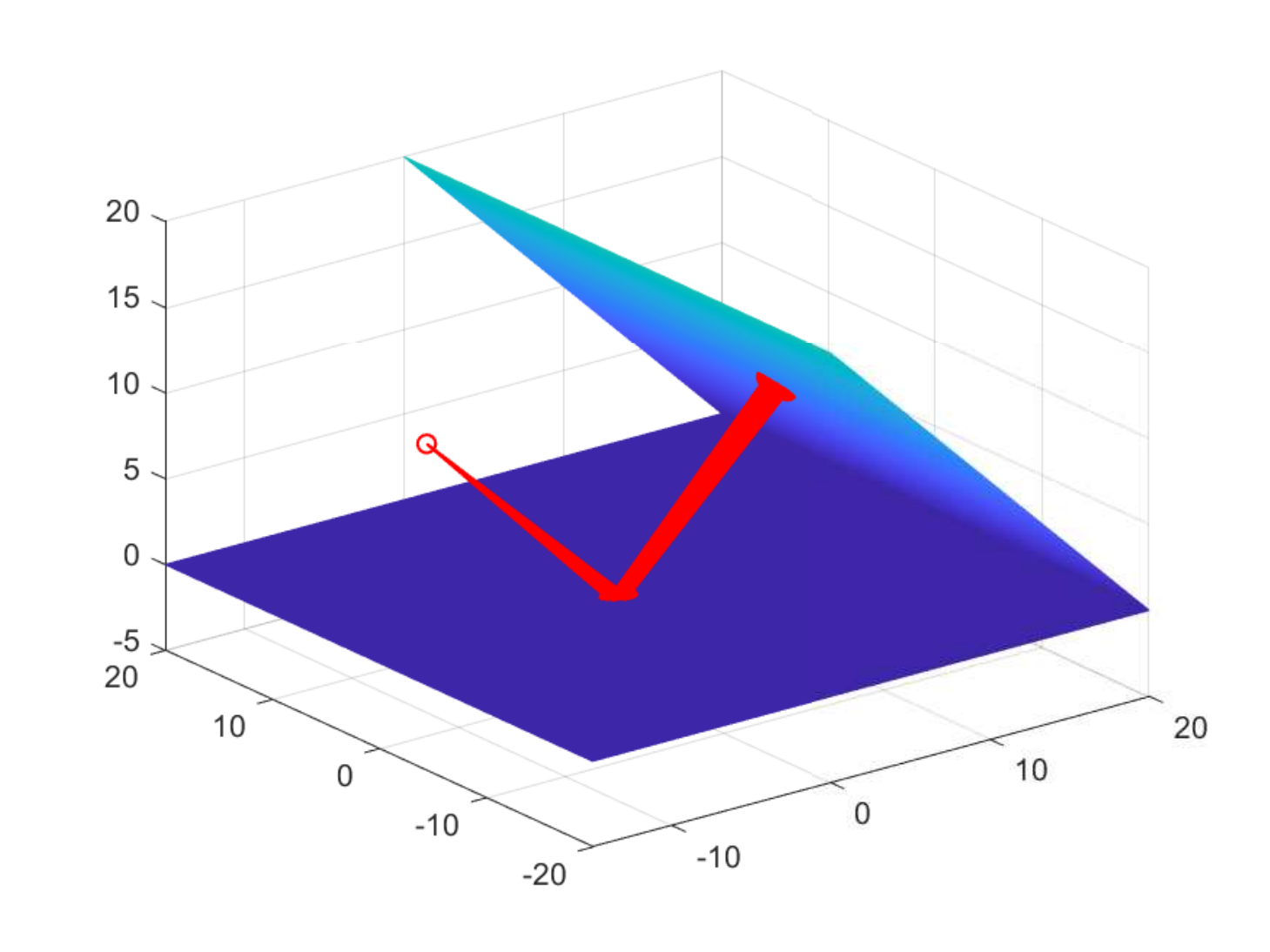}
\caption{Simulation of the optical path with beam and ICRN jitter ($\sigma_{\theta_k}=5\times10^{-2}$, $\sigma_{\beta_k}=1\times10^{-2}$).}
\label{fig.5}
\end{figure}

\begin{figure}[htbp]
\centering
\includegraphics[width=1\textwidth]{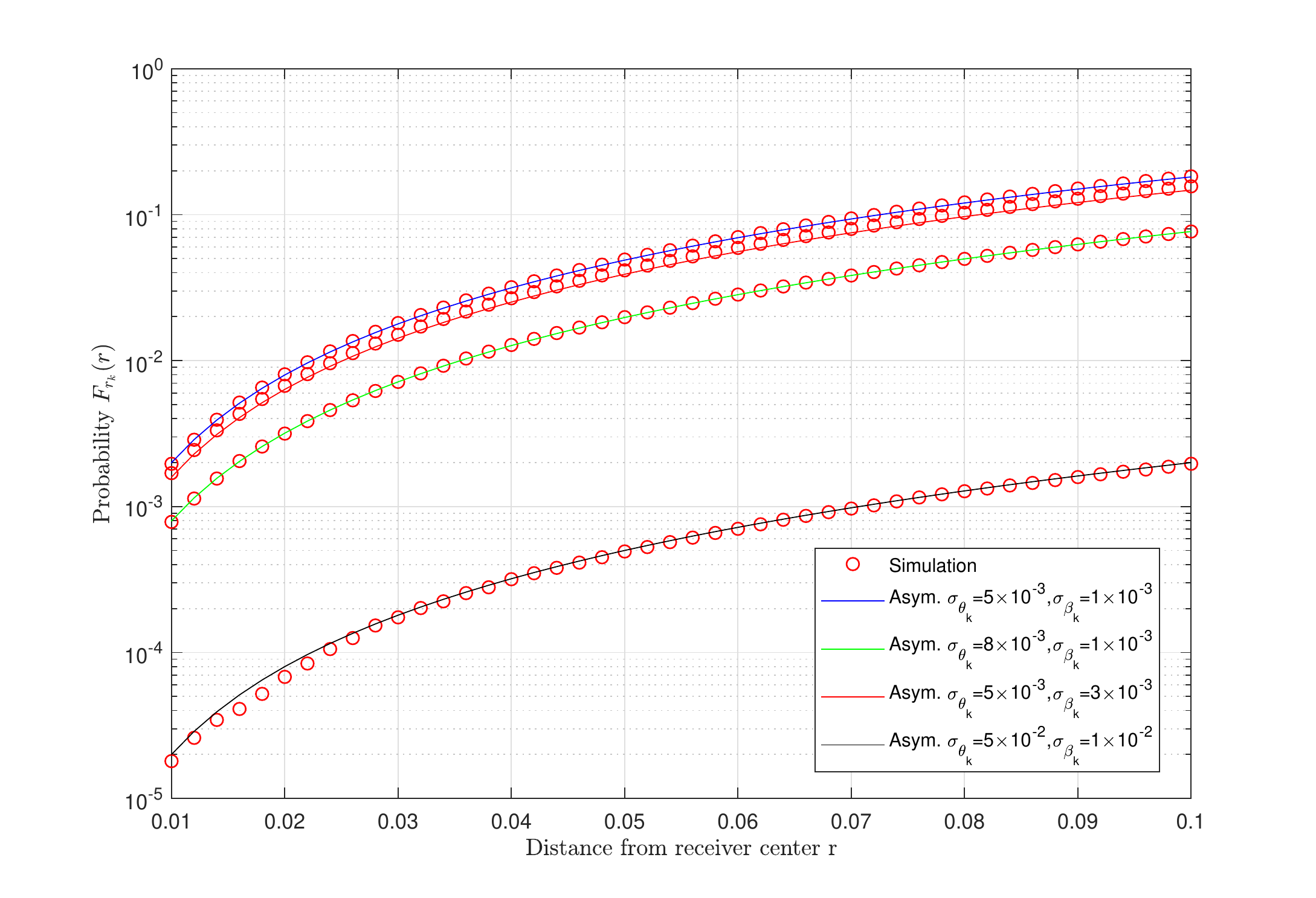}
\caption{The asymptotic CDF and simulated CDF for beam offset at the receiver ($w_k=4\sqrt{3}, l_k=2\sqrt{10}$), the asymptotic results are obtained from \eqref{8-1}.}
\label{fig.6}
\end{figure}

In Fig. \ref{fig.5}, the optical path with beam and ICRN jitter is simulated. Twenty million sets of noise is added to the beam direction at the transmitting end and ICRN normal vector to simulate the actual jitter. It can be intuitively seen from Fig. \ref{fig.5} that the jitter of the outgoing beam at the transmitting end is amplified after being reflected by the ICRN plane.
In Fig. \ref{fig.6}, we respectively present the asymptotic CDF and computer simulated CDF for beam offset at the receiver. Monte Carlo method is used in the simulation to estimate the CDF of the beam offset by counting the number of points in the receiving plane at different distances from the center of the receiver. The CDF of $r_k$ can be derived from \eqref{3-7} and \eqref{5-1} as
\begin{equation} \label{8-1}
\begin{aligned}
F_{r_k}(r)&=P(r_k\leq r)=F_{\theta_{k}^{(s)}}(\frac{r}{l_k})\\&=1-exp\left(\frac{-r^2}{2\left(l_k+w_k\right)^2\sigma _{\theta _{k}}^{2}+  8\sigma _{\beta _{k}}^{2}l_{k}^{2}}\right).
\end{aligned}
\end{equation}
It can be seen from Fig. \ref{fig.6} that the asymptotic curve agrees well with the simulation results. An increase in the standard deviation of pointing error angle $\sigma_{\theta_k}$ and deflection error angle $\sigma_{\beta_k}$ leads to a decrease in $F_{r_k}(r)$, which is consistent with \eqref{8-1}.

\subsection{Numerical Simulation of System Performance}
\begin{figure}[htbp]
\centering
\includegraphics[width=1\textwidth]{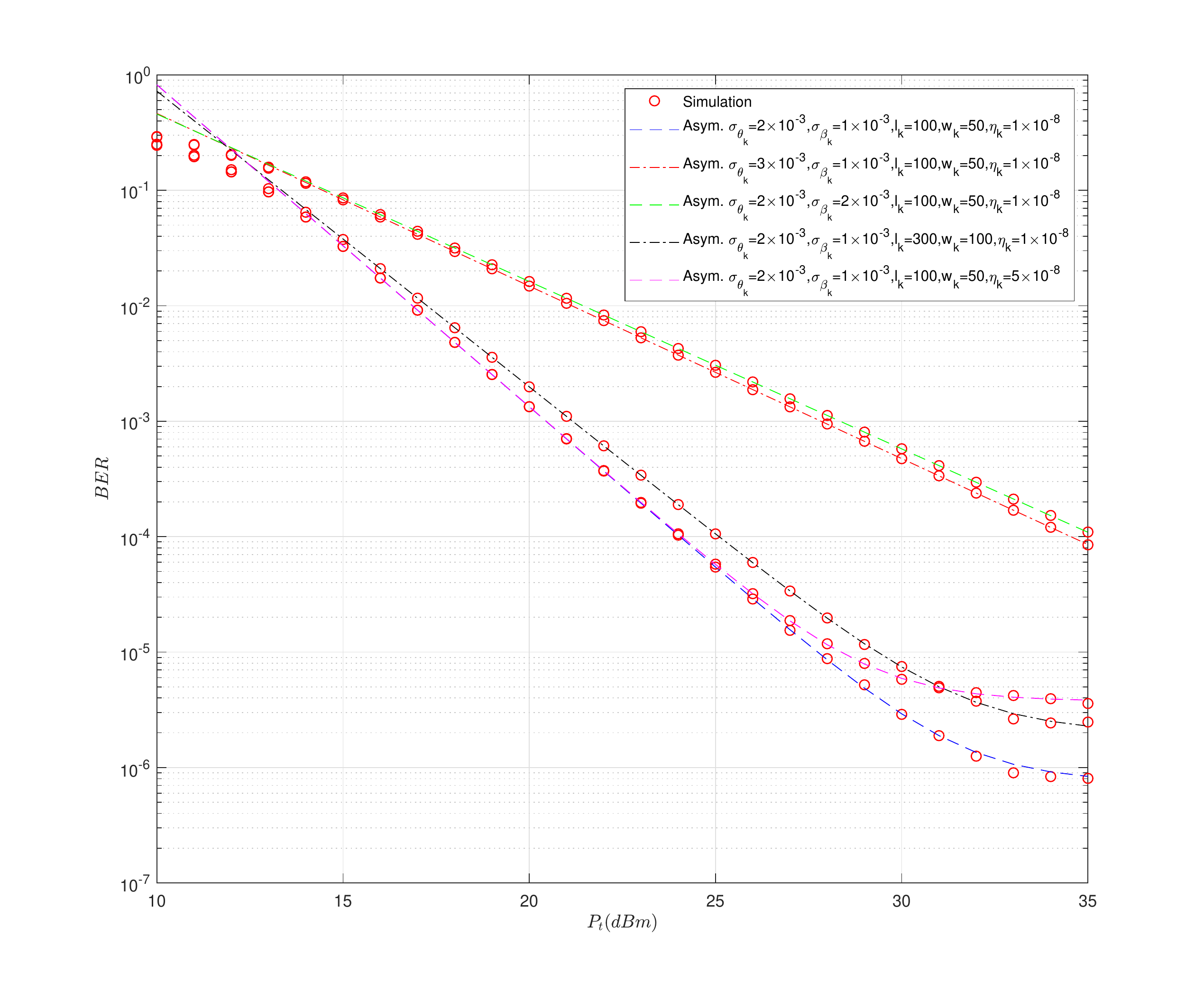}
\caption{The asymptotic BERs and computer simulated BERs for single intelligent channel ($\sigma_n=10^{-2}, \phi=8\times10^{-3}rad$) with different jitter and obstruction probability values, the asymptotic results are obtained from \eqref{7-6-2}.}
\label{fig.7}
\end{figure}

\begin{figure}[htbp]
\centering
\includegraphics[width=1\textwidth]{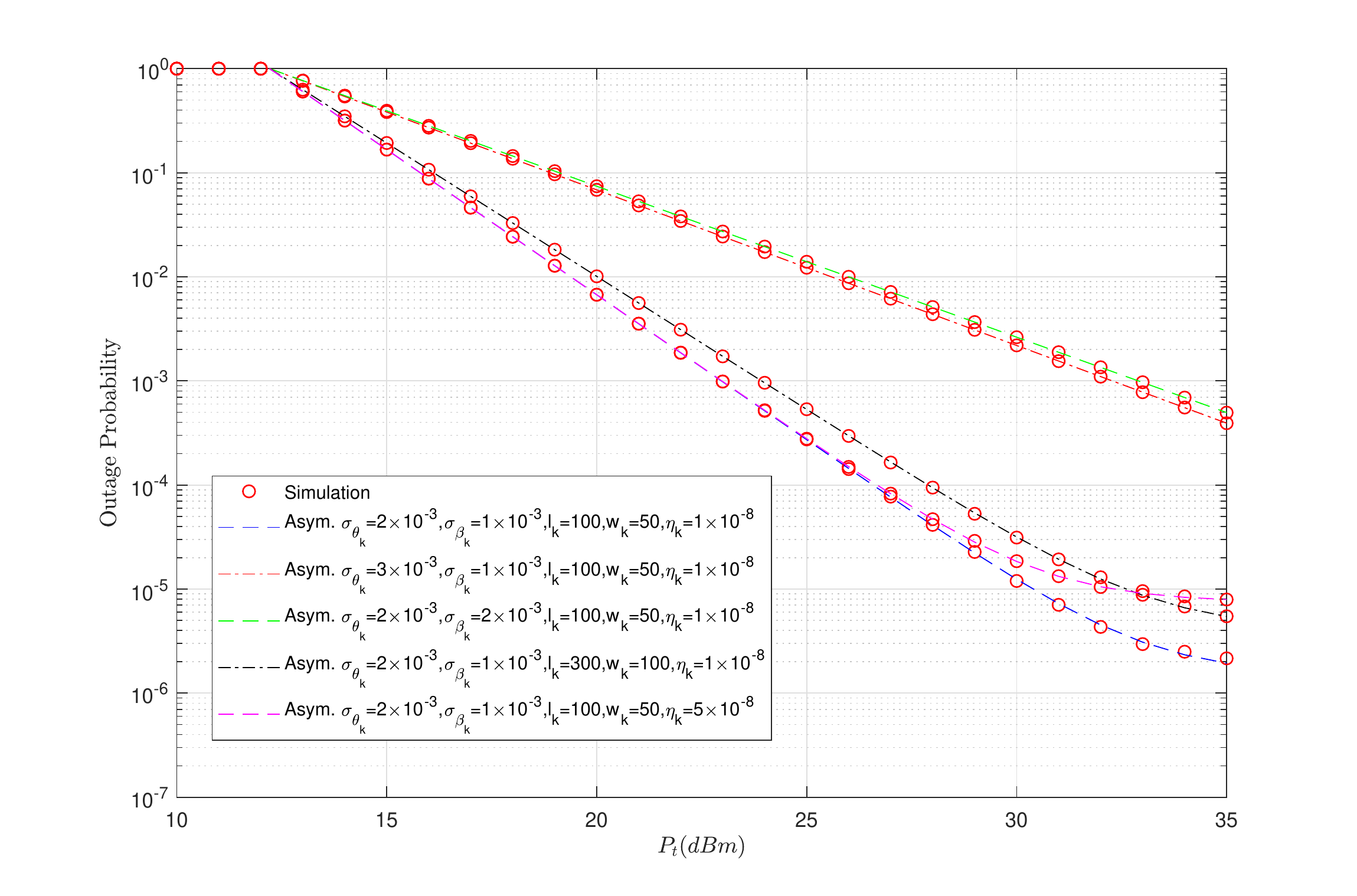}
\caption{Outage probability for single intelligent channel ($\sigma_n=10^{-2}, \phi=8\times10^{-3}rad$) with different jitter and obstruction probability values, the asymptotic results are obtained from \eqref{7-7}.}
\label{fig.8}
\end{figure}

In Fig. \ref{fig.7}, we respectively present asymptotic BERs and simulated BERs for single intelligent channel with different jitter values. The asymptotic BER curves are obtained by \eqref{7-6}. The outage probability curves for the same systems with SNR threshold $\gamma_{th}=5\ dB$ are presented in Fig. \ref{fig.8}, where the asymptotic outage probability curves are obtained by \eqref{7-7}. From Fig. \ref{fig.7}, we observe that the simulated BER curves for IM/DD with OOK modulation agree well with the asymptotic BER curves in high SNR regimes. From Fig. \ref{fig.8}, the same behavior can be observed for outage probability. The numerical results indicate that the asymptotic estimation of
system performance measures is accurate in large SNR regimes. Figs. \ref{fig.7} and \ref{fig.8} show that there exists BER and outage probability floor at high SNR, which is caused by the probability of obstruction.  This observation is expected because from \eqref{7-6} and \eqref{7-7} we can derive the lower bounds of BER and outage probability. From Figs. \ref{fig.7} and \ref{fig.8}, we can observe that an increase of $\sigma_{\theta_k}, \sigma_{\beta_k}, l_k, w_k, \eta_k$ will all lead to a decrease in system performance, where $\eta_k, w_k$ and $l_k$  affect the BER and outage probability level, and $\sigma_{\theta_k}, \sigma_{\beta_k}$ affect the convergence speed of BER and outage probability.\par
\begin{figure}[htbp]
\centering
\includegraphics[width=1\textwidth]{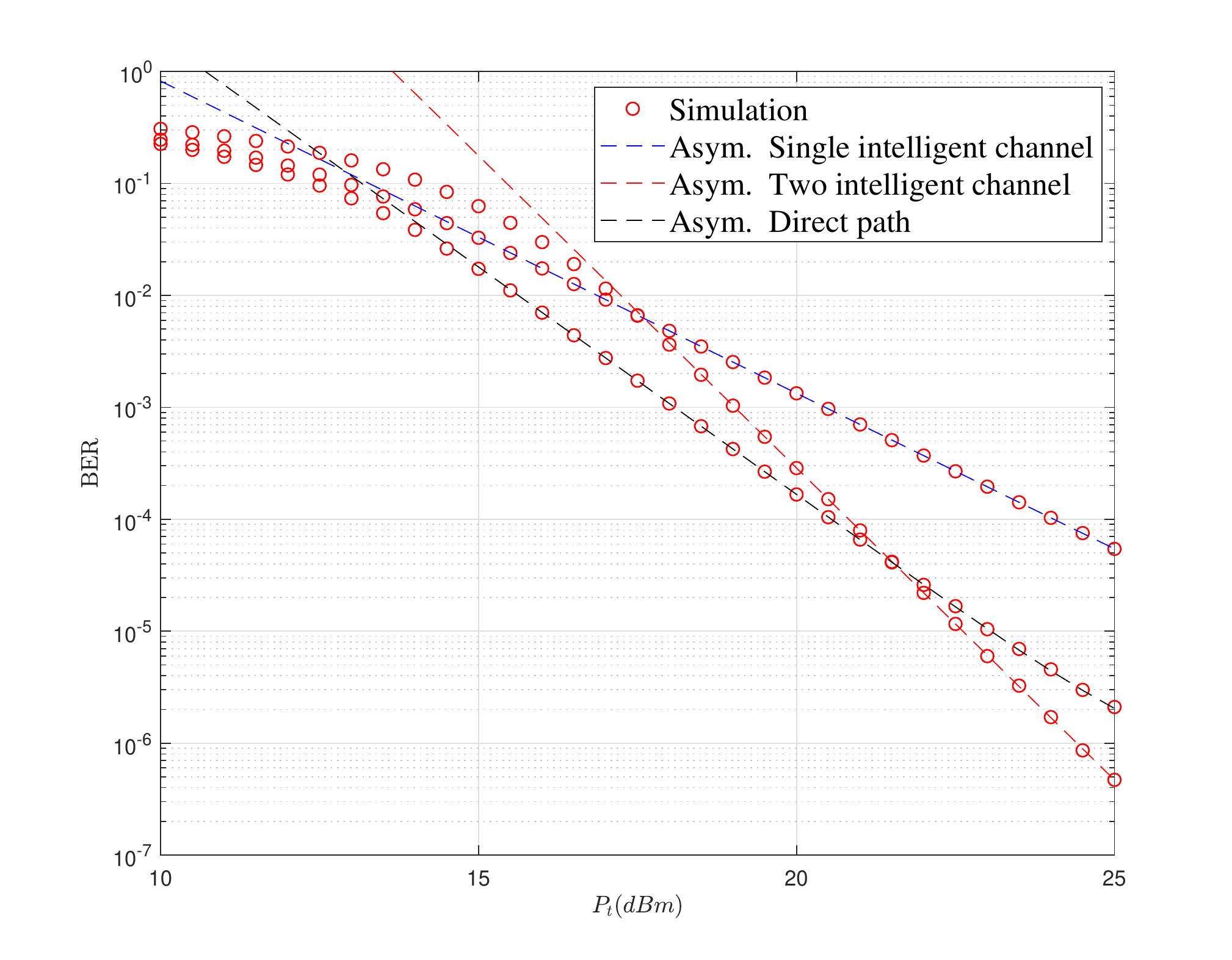}
\caption{Comparison of BER for systems of single intelligent channel, two intelligent channels and direct path (parameters of the intelligent channel and direct path are shown in Table \ref{table.1}), the asymptotic results are obtained from \eqref{7-6-2} and \eqref{7-13}.}
\label{fig.9}
\end{figure}

\begin{figure}[htbp]
\centering
\includegraphics[width=1\textwidth]{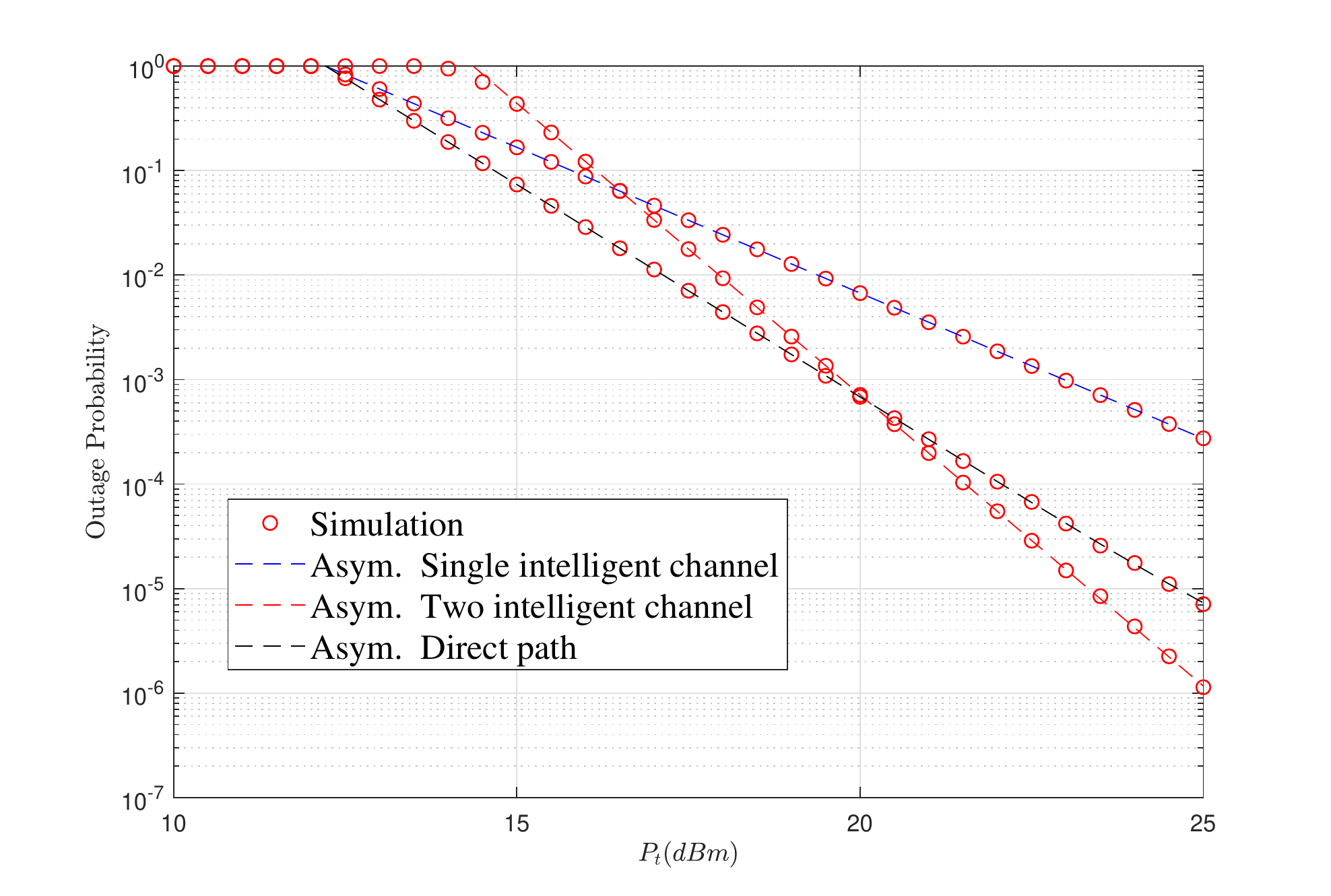}
\caption{Comparison of outage probability for systems of single intelligent channel, two intelligent channels and direct path (parameters of the intelligent channel and direct path are shown in Table \ref{table.1}), the asymptotic results are obtained from \eqref{7-7} and \eqref{7-14}.}
\label{fig.10}
\end{figure}

Figs. \ref{fig.9} and \ref{fig.10} show the comparison of BER and outage probability with $\gamma_{th}=5\ dB$ for systems of single intelligent channel, two intelligent channels and direct path. The asymptotic results of FSO system with direct path can be obtained by \cite{uysal2006error}\cite{borah2009pointing}. For comparison, we suppose all the intelligent channels in the systems are the same in $\sigma_{\theta_k}, \sigma_{\beta_k}, l_k, w_k$, and $\eta_k$. The parameters in the systems are presented in Table \ref{table.1}. From Figs. \ref{fig.9} and \ref{fig.10}, we can observe that the simulated curves agree well with asymptotic curves at high SNR, which indicates that the asymptotic estimation is accurate. Figs. \ref{fig.9} and \ref{fig.10} show that the system of direct path has better BER and outage probability performance than the system of single intelligent channel. However, the system of two intelligent channels has the best performance and the lowest BER and outage probability floor among the three systems, which indicates that adding one intelligent channel significantly improves system performance.
\begin{table}[!htbp]
\centering
\caption{SYSTEM SETTINGS}\label{}
\begin{tabular}{c|c}
\hline
Parameters of the intelligent channel& value\\
\hline
Receiver Diameter (2a)& 20 cm\\
Noise variance ($\sigma_n^2$)& $10^{-4}$ W\\
Link distance from transmitter to ICRN ($w_k$)& 50 m\\
Link distance from ICRN to receiver ($l_k$)& 100 m \\
Transmit Divergence at $1/e^2$ ($\phi$)& 8 mrad \\
Corresponding beam radius ($w_z$)& $\approx$ 120 cm\\
Pointing error angle standard deviation ($\sigma_{\theta}$)& 5 mrad \\
ICRN jitter angle standard deviation ($\sigma_{\beta}$)& 2 mrad \\
Obstacle probability coefficient $(\eta_k)$ & $10^{-8}$\\
\hline
Parameters of the channel with direct path& value\\
\hline
Receiver Diameter (2a)& 20 cm\\
Noise variance ($\sigma_n^2$)& $10^{-4}$ W\\
Link distance from transmitter to receiver & 100 m\\
Transmit Divergence at $1/e^2$ ($\phi$)& 8 mrad \\
Corresponding beam radius ($w_z$)& $\approx$ 80 cm\\
Pointing error angle standard deviation ($\sigma_{\theta}$)& 5 mrad \\
Obstacle probability coefficient $(\eta_k)$ & $10^{-8}$ \\
\hline
\label{table.1}
\end{tabular}

\end{table}

\begin{figure}[htbp]
\centering
\includegraphics[width=1\textwidth]{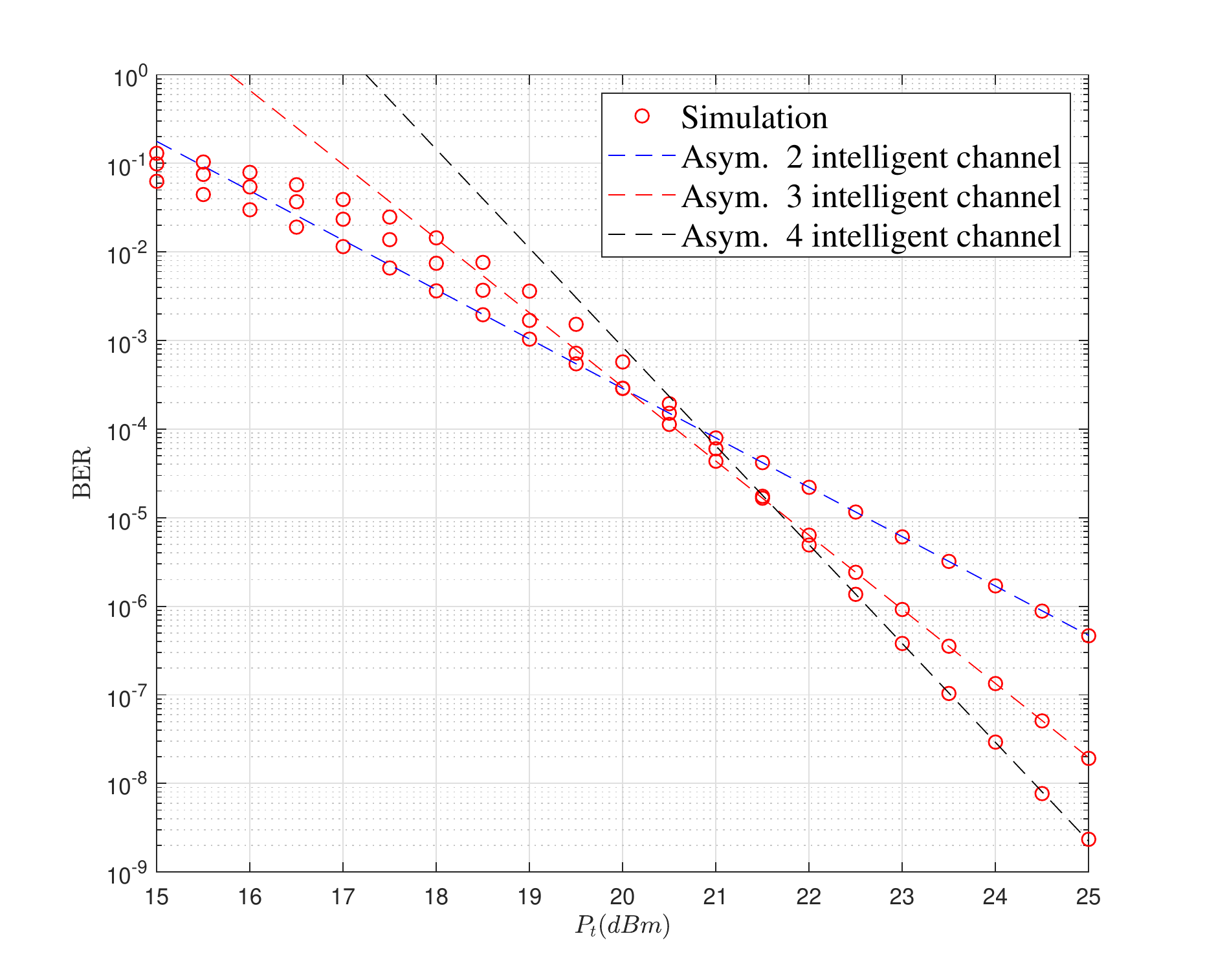}
\caption{BER for systems with different number of intelligent channels (parameters of the intelligent channel are shown in Table \ref{table.1}), the asymptotic results are obtained from \eqref{7-13}.}
\label{fig.11}
\end{figure}

\begin{figure}[htbp]
\centering
\includegraphics[width=1\textwidth]{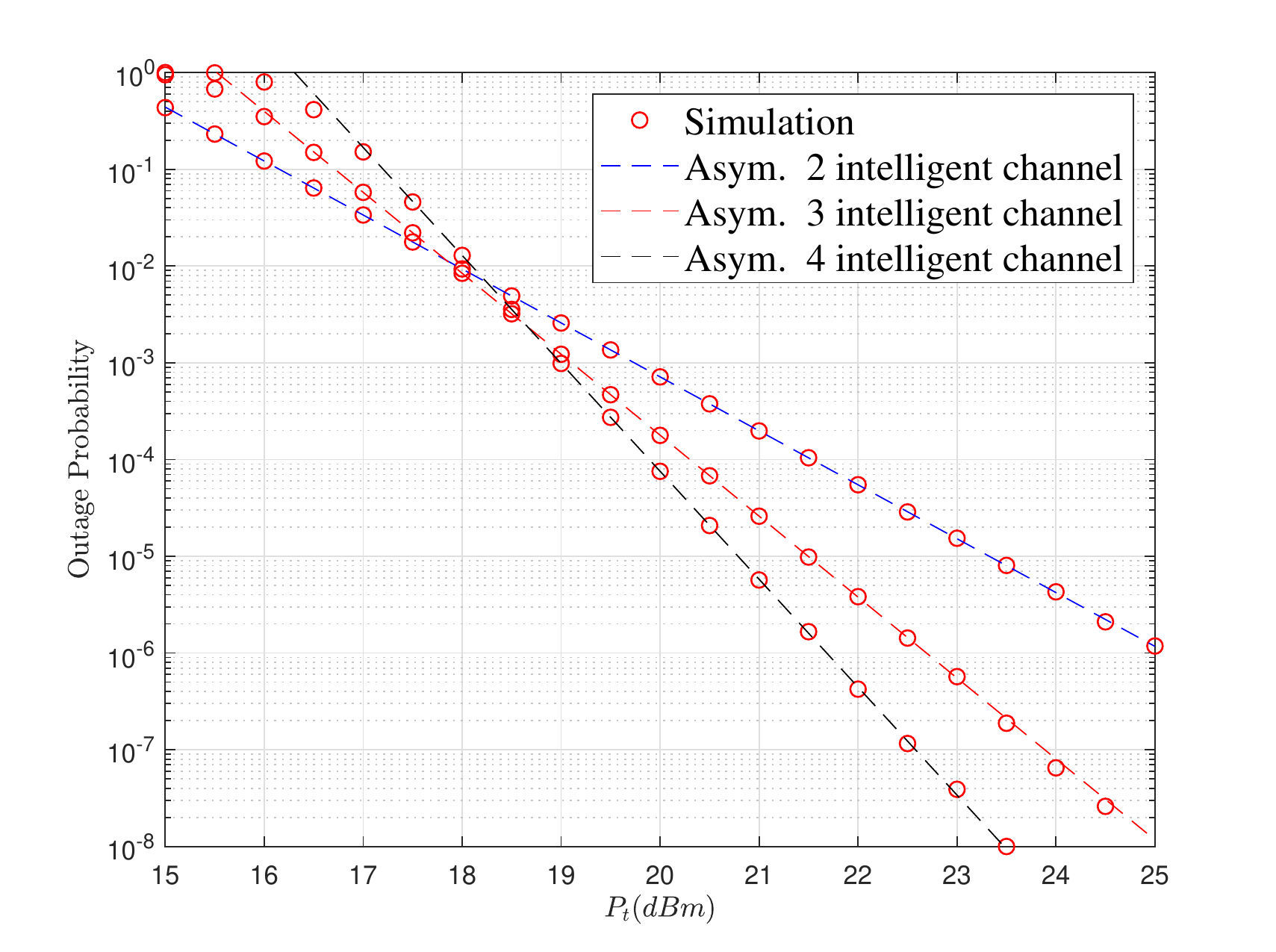}
\caption{Outage probability for systems with different number of intelligent channels (parameters of the intelligent channel are shown in Table \ref{table.1}), the asymptotic results are obtained from \eqref{7-14}.}
\label{fig.12}
\end{figure}

\begin{figure}[htbp]
\centering
\includegraphics[width=1\textwidth]{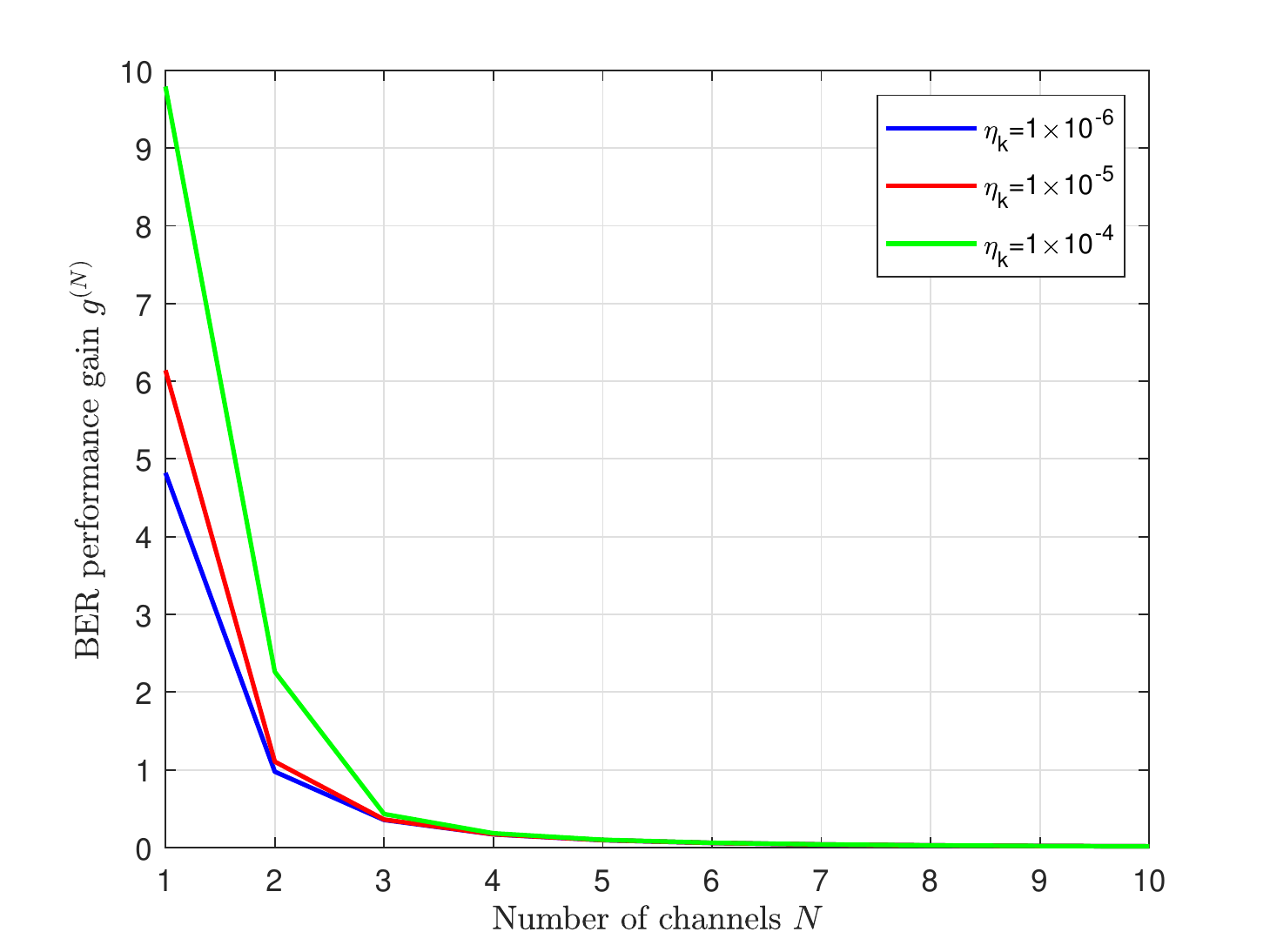}
\caption{The relationship between the BER performance gain $g^{(N)}$ and the number of channels $N$ at $P_t=20\ dBm$ (parameters of the intelligent channel are shown in Table \ref{table.1}), the asymptotic results are obtained from \eqref{9-2}.}
\label{fig.13}
\end{figure}

In Figs. \ref{fig.11} and \ref{fig.12}, we compare the BER and outage probability with $\gamma_{th}=5\ dB$ for systems of different number of intelligent channels. For comparison, we assume that the parameters of all intelligent channels in the system are exactly the same. We can observe the asymptotic results agree well with the simulation results at high SNR, which indicates that the asymptotic estimation is accurate. Figs. \ref{fig.11} and \ref{fig.12} show that the performance gap between systems of 2 channels and 3 channels is larger than that between 3 channels and 4 channels, which indicates that as the number of channels increases, the performance gain by adding one channel becomes smaller.
Fig. \ref{fig.13} shows the relationship between the BER performance gain $g^{(N)}$ and the number of channels $N$ in the system at $P_t=20\ dBm$. From Fig.\ref{fig.13}, we can observe that as the number of channels in the system increases, the BER gain brought by adding an intelligent channel decreases continuously. Therefore, when designing the system, we do not need to blindly increase the number of intelligent channels to improve the performance of the system. Fig. \ref{fig.13} shows that as the probability of obstruction increases, the BER performance gain brought by the increase in the number of channels becomes greater, which indicates that adding intelligent channels to the system is an effective method to deal with the obstruction of obstacles that may appear in the channel.

\section{Conclusion}\label{conclusion}
In this paper, we change the propagation path of optical signals by adding optical RIS to the free space optical channel and use multiple optical RIS to implement the diversity transmission of optical signals. In the channel modeling, the influence of the reflection of the optical path on the pointing error and channel fading, the influence of the RIS surface jitter, and the probability of obstacles in the channel are taken into consideration and investigated. Based on the asymptotic analysis and computer simulation, we observe that using optical RIS to increase the number of controllable channels can effectively improve system performance and reduce the probability of communication systems being interrupted by obstacles in the environment. However, as the number of channels increases, the performance gain caused by increasing the number of channels continues to decrease. Therefore, we need to determine the number and placement of RIS according to the actual situation, so as to achieve better communication performance with a lower cost. In addition, in this paper, we set one ICRN to each intelligent channel and suppose each channel is independent of each other, where has no case of signal multi-hop transmission between ICRNs in multiple channels. In subsequent investigation, the effect of multi-hop transmission of signals between ICRNs on system performance can be further studied and the effect of signal multi-hop reflection on the pointing error can be further deduced.

\begin{appendices}
\section{Derivation of $\theta_{x_k}^{(s)}$} \label{appA}
This is the derivation process of \eqref{3-3-1}. In Fig.\ref{fig.3}, $O$ is the intersection of desired aiming light beam and original ICRN plane, and $O^{(x)}$ is the intersection of horizontal component of actual light beam with jitter and original ICRN plane, where the position of the reflection point of the actual light beam is assumed to be in the original ICRN plane to simplify the expression of $r_{k}$. The error term of $r_{k}$ by this assumption is $\frac{\theta_{x_k}\beta_{x_k}w_k}{cos(\alpha+\theta_{x_k}+\beta_{x_k})}$, which can be discarded when $\theta_{x_k}$ and $\beta_{x_k}$ are small, where $\alpha$ is incidence angle of desired aiming light beam. The $B_x$ is the intersection of desired aiming light beam and receiver plane and $C_x$ is the intersection of actual light beam with jitter and receiver plane. The extension lines of $B_xO$ and $C_xO^{(x)}$ intersect at point $D_x$.\par
$\because \angle D_xOO^{(x)}=90^{\circ}+\alpha , \angle D_xO^{(x)}O=180^{\circ}-\angle AO^{(x)}C_x-\angle AO^{(x)}O=90^{\circ}-\alpha-2\beta_{x_k}-\theta_{x_k}$\par
$\therefore \angle OD_xO^{(x)}=180^{\circ}-\angle D_xOO^{(x)}-\angle D_xO^{(x)}O=2\beta_{x_k}+\theta_{x_k}$\par
$\because tan\angle OC_xO^{(x)}\cdot l_k=OO^{(x)}\cdot sin\angle D_xO^{(x)}O=OO^{(x)}\cdot cos(\alpha+2\beta_{x_k}+\theta_{x_k})$,\par
$\quad tan\theta_{x_k}\cdot w_k=OO^{(x)}\cdot sin\angle AO^{(x)}O=OO^{(x)}\cdot cos(\alpha+\theta_{x_k})$, \par
$\quad \theta_{x_k}\ and\ \beta_{x_k}$ are small compared with incidence angle $\alpha$\par
$\therefore \theta_{x_k}\cdot w_k \approx \angle OC_xO^{(x)}\cdot l_k$\par
$\therefore \angle OC_xO^{(x)}\approx\frac{\theta_{x_k}\cdot w_k}{l_k}$\par
$\therefore \theta_{x_k}^{(s)}=\angle OD_xO^{(x)}+\angle OC_xO^{(x)}\approx (1+\frac{w_k}{l_k})\theta_{x_k}+2\beta_{x_k}$
\section{Derivation of $\theta_{y_k}^{(s)}$} \label{appB}
This is the derivation process of \eqref{3-3-2}. In Fig.\ref{fig.4}, $O$ is the intersection of desired aiming light beam and original ICRN plane, and $O^{(y)}$ is the intersection of vertical component of actual light beam with jitter and original ICRN plane, where the position of the reflection point of the actual light beam is assumed to be in the original ICRN plane to simplify the expression of $r_{k}$. The error term of $r_{k}$ by this assumption is $\frac{\theta_{y_k}\beta_{y_k}w_k}{cos(\theta_{y_k}+\beta_{y_k})}$, which can be discarded when $\theta_{y_k}$ and $\beta_{y_k}$ are small. The $B_y$ is the intersection of desired aiming light beam and receiver plane and $C_y$ is the intersection of vertical component of actual light beam with jitter and receiver plane. The extension lines of $B_yO$ and $C_yO^{(y)}$ intersect at point $D_y$.\par
$\because \angle D_yOO^{(y)}=90^{\circ} , \angle D_yO^{(y)}O=180^{\circ}-\angle AO^{(y)}C_y-\angle AO^{(y)}O=90^{\circ}-2\beta_{y_k}-\theta_{y_k}$\par
$\therefore \angle OD_yO^{(y)}=180^{\circ}-\angle D_yOO^{(y)}-\angle D_yO^{(y)}O=2\beta_{y_k}+\theta_{y_k}$\par
$\because tan\angle OC_yO^{(y)}\cdot l_k=OO^{(y)}\cdot sin\angle D_yO^{(y)}O=OO^{(y)}\cdot cos(2\beta_{y_k}+\theta_{y_k})$,\par
$\quad tan\theta_{y_k}\cdot w_k=OO^{(y)}\cdot sin\angle AO^{(y)}O=OO^{(y)}\cdot cos\theta_{y_k}$, \par
$\quad \theta_{y_k}\ and\ \beta_{y_k}$ are small\par
$\therefore \theta_{y_k}\cdot y_k \approx \angle OC_yO^{(y)}\cdot l_k$\par
$\therefore \angle OC_yO^{(y)}\approx\frac{\theta_{y_k}\cdot w_k}{l_k}$\par
$\therefore \theta_{y_k}^{(s)}=\angle OD_xO^{(y)}+\angle OC_yO^{(y)}\approx (1+\frac{w_k}{l_k})\theta_{y_k}+2\beta_{y_k}$
\section{Power Allocation Schenme} \label{PowerAllocationProof}
This is the solution process of \eqref{d-1}. Firstly, we can obtain the Lagrange function according to the objective function and constraints as
\begin{equation} \label{a-1}
\begin{aligned}
L(\alpha_0,\alpha_1,\cdots,\alpha_{M-1})&=\frac{1}{2}\prod_{k=0}^{M-1}(1-n_k)+\sum_{k=0}^{M-1}m_kn_k2^{m_k-1}\left(\frac{\sigma_n^2}{2P_t^2A_0^2\alpha_k^2}\right)^{\frac{m_k}{2}}\Gamma(\frac{m_k+1}{2})\frac{\prod_{i=0}^{M-1}(1-n_i)}{\sqrt{\pi}(1-n_k)m_k}
\\&+\lambda (\sum_{k=0}^{M-1}\alpha_k-1)
\end{aligned}
\end{equation}
where $\lambda$ is the Lagrange multiplier. By taking the partial derivatives of the $L(\alpha_0,\alpha_1,\cdots,\alpha_{M-1})$ with respect to $\alpha_0,\alpha_1,\cdots,\alpha_{M-1}$ and $\lambda$, we can obtain the equation set as
\begin{equation} \label{a-2}
\begin{aligned}
\left\{\begin{matrix}m_0b_0\alpha_0^{-m_0-1}-\lambda =0
\\
m_1b_1\alpha_1^{-m_1-1}-\lambda =0
\\
\vdots
\\
m_ib_i\alpha_i^{-m_i-1}-\lambda =0
\\
\vdots
\\
m_{M-1}b_{M-1}\alpha_{M-1}^{-m_{M-1}-1}-\lambda =0
\\
\sum_{i=0}^{M-1}\alpha_i-1=0
\end{matrix}\right.
\end{aligned}
\end{equation}
where
\begin{equation} \label{a-3}
\begin{aligned}
b_i=m_in_i2^{m_i-1}\left(\frac{\sigma_n^2}{2P_t^2A_0^2}\right)^{\frac{m_i}{2}}\Gamma(\frac{m_i+1}{2})\frac{\prod_{j=0}^{M-1}(1-n_j)}{\sqrt{\pi}(1-n_i)m_i}, \quad\quad i=0,1,2,\cdots,M-1.
\end{aligned}
\end{equation}
The solution can be obtained by solving the equation set as
\begin{equation} \label{a-4}
\begin{aligned}
\alpha_i=\frac{(b_im_i)^{\frac{1}{m_i+1}}}{\sum_{j=0}^{M-1}(b_jm_j)^{\frac{1}{m_j+1}}}, \quad\quad i=0,1,2,\cdots,M-1.
\end{aligned}
\end{equation}
\end{appendices}
\bibliographystyle{IEEEtran}
\bibliography{IEEEabrv,ref}

\begin{thebibliography}{10}
\providecommand{\url}[1]{#1}
\csname url@samestyle\endcsname
\providecommand{\newblock}{\relax}
\providecommand{\bibinfo}[2]{#2}
\providecommand{\BIBentrySTDinterwordspacing}{\spaceskip=0pt\relax}
\providecommand{\BIBentryALTinterwordstretchfactor}{4}
\providecommand{\BIBentryALTinterwordspacing}{\spaceskip=\fontdimen2\font plus
\BIBentryALTinterwordstretchfactor\fontdimen3\font minus
  \fontdimen4\font\relax}
\providecommand{\BIBforeignlanguage}[2]{{%
\expandafter\ifx\csname l@#1\endcsname\relax
\typeout{** WARNING: IEEEtran.bst: No hyphenation pattern has been}%
\typeout{** loaded for the language `#1'. Using the pattern for}%
\typeout{** the default language instead.}%
\else
\language=\csname l@#1\endcsname
\fi
#2}}
\providecommand{\BIBdecl}{\relax}
\BIBdecl

\bibitem{lin2019millimeter}
Z.~Lin, X.~Du, H.-H. Chen, B.~Ai, Z.~Chen, and D.~Wu, ``Millimeter-wave
  propagation modeling and measurements for 5g mobile networks,'' \emph{IEEE
  Wirel Commun.}, vol.~26, no.~1, pp. 72--77, 2019.

\bibitem{ni2019research}
Y.~Ni, J.~Liang, X.~Shi, and D.~Ban, ``Research on key technology in 5g mobile
  communication network,'' in \emph{2019 International Conference on
  Intelligent Transportation, Big Data \& Smart City (ICITBS)}.\hskip 1em plus
  0.5em minus 0.4em\relax IEEE, 2019, pp. 199--201.

\bibitem{nishimura2019optical}
K.~Nishimura, S.~Ishimura, A.~Bekkali, K.~Tanaka, H.~Hirayama, Y.~Tsukamoto,
  S.~Nanba, and M.~Suzuki, ``Optical access technology for b5g mfh/mbh,'' in
  \emph{Optical Fiber Communication Conference}.\hskip 1em plus 0.5em minus
  0.4em\relax Optical Society of America, 2019, pp. W3J--1.

\bibitem{xiao2019resource}
Y.~Xiao, J.~Zhang, Z.~Liu, and Y.~Ji, ``Resource-efficient slicing for 5g/b5g
  converged optical-wireless access networks,'' in \emph{Asia Communications
  and Photonics Conference}.\hskip 1em plus 0.5em minus 0.4em\relax Optical
  Society of America, 2019, pp. M4A--202.

\bibitem{busari2019terahertz}
S.~A. Busari, K.~M.~S. Huq, S.~Mumtaz, and J.~Rodriguez, ``Terahertz massive
  mimo for beyond-5g wireless communication,'' in \emph{ICC 2019-2019 IEEE
  International Conference on Communications (ICC)}.\hskip 1em plus 0.5em minus
  0.4em\relax IEEE, 2019, pp. 1--6.

\bibitem{song2011present}
H.-J. Song and T.~Nagatsuma, ``Present and future of terahertz
  communications,'' \emph{IEEE Trans.Thz Sci Technol.}, vol.~1, no.~1, pp.
  256--263, 2011.

\bibitem{rebeiz1992millimeter}
G.~M. Rebeiz, ``Millimeter-wave and terahertz integrated circuit antennas,''
  \emph{P. IEEE}, vol.~80, no.~11, pp. 1748--1770, 1992.

\bibitem{siegel2002terahertz}
P.~H. Siegel, ``Terahertz technology,'' \emph{IEEE Trans. Microw Theory},
  vol.~50, no.~3, pp. 910--928, 2002.

\bibitem{rappaport2014mobile}
T.~S. Rappaport, W.~Roh, and K.~Cheun, ``Mobile's millimeter-wave makeover,''
  \emph{IEEE Spectr.}, vol.~51, no.~9, pp. 34--58, 2014.

\bibitem{huang2019reconfigurable}
C.~Huang, A.~Zappone, G.~C. Alexandropoulos, M.~Debbah, and C.~Yuen,
  ``Reconfigurable intelligent surfaces for energy efficiency in wireless
  communication,'' \emph{IEEE Trans. Wireless Commun.}, vol.~18, no.~8, pp.
  4157--4170, 2019.

\bibitem{li2019reconfigurable}
S.~Li, B.~Duo, X.~Yuan, Y.-C. Liang, M.~Di~Renzo \emph{et~al.},
  ``Reconfigurable intelligent surface assisted uav communication: Joint
  trajectory design and passive beamforming,'' \emph{arXiv preprint
  arXiv:1908.04082}, 2019.

\bibitem{di2019hybrid}
B.~Di, H.~Zhang, L.~Song, Y.~Li, Z.~Han, and H.~V. Poor, ``Hybrid beamforming
  for reconfigurable intelligent surface based multi-user communications:
  Achievable rates with limited discrete phase shifts,'' \emph{arXiv preprint
  arXiv:1910.14328}, 2019.

\bibitem{wu2019intelligent}
Q.~Wu and R.~Zhang, ``Intelligent reflecting surface enhanced wireless network
  via joint active and passive beamforming,'' \emph{IEEE Trans. Wireless
  Commun.}, vol.~18, no.~11, pp. 5394--5409, 2019.

\bibitem{vidal2006optical}
B.~Vidal, T.~Mengual, C.~Ibanez-Lopez, and J.~Marti, ``Optical beamforming
  network based on fiber-optical delay lines and spatial light modulators for
  large antenna arrays,'' \emph{IEEE Photon. Technol. Lett.}, vol.~18, no.~24,
  pp. 2590--2592, 2006.

\bibitem{armitage1985high}
D.~Armitage, W.~Anderson, and T.~Karr, ``High-speed spatial light modulator,''
  \emph{IEEE J. Quantum Electron.}, vol.~21, no.~8, pp. 1241--1248, 1985.

\bibitem{ross1982two}
W.~E. Ross, D.~Psaltis, and R.~H. Anderson, ``Two-dimensional magneto-optic
  spatial light modulator for signal processing,'' in \emph{Real-Time Signal
  Processing V}, vol. 341.\hskip 1em plus 0.5em minus 0.4em\relax International
  Society for Optics and Photonics, 1982, pp. 191--198.

\bibitem{ma2003optical}
X.~Ma and G.-S. Kuo, ``Optical switching technology comparison: optical mems
  vs. other technologies,'' \emph{IEEE Commun. Mag.}, vol.~41, no.~11, pp.
  S16--S23, 2003.

\bibitem{kim2013wireless}
S.-M. Kim and S.-M. Kim, ``Wireless visible light communication technology
  using optical beamforming,'' \emph{Optical Engineering}, vol.~52, no.~10, p.
  106101, 2013.

\bibitem{zhang2018optical}
Z.~Zhang, J.~Dang, L.~Wu, H.~Wang, J.~Xia, W.~Lei, J.~Wang, and X.~You,
  ``Optical mobile communications: Principles, implementation, and performance
  analysis,'' \emph{IEEE Trans. Veh. Technol.}, vol.~68, no.~1, pp. 471--482,
  2018.

\bibitem{sandalidis2008ber}
H.~G. Sandalidis, T.~A. Tsiftsis, G.~K. Karagiannidis, and M.~Uysal, ``Ber
  performance of fso links over strong atmospheric turbulence channels with
  pointing errors,'' \emph{IEEE Commun. Lett.}, vol.~12, no.~1, pp. 44--46,
  2008.

\bibitem{garcia2009selection}
A.~Garcia-Zambrana, C.~Castillo-Vazquez, B.~Castillo-Vazquez, and
  A.~Hiniesta-Gomez, ``Selection transmit diversity for fso links over strong
  atmospheric turbulence channels,'' \emph{IEEE Photon. Technol. Lett.},
  vol.~21, no.~14, pp. 1017--1019, 2009.

\bibitem{uysal2006error}
M.~Uysal, J.~Li, and M.~Yu, ``Error rate performance analysis of coded
  free-space optical links over gamma-gamma atmospheric turbulence channels,''
  \emph{IEEE Trans. Wireless Commun.}, vol.~5, no.~6, pp. 1229--1233, 2006.

\bibitem{borah2009pointing}
D.~K. Borah and D.~G. Voelz, ``Pointing error effects on free-space optical
  communication links in the presence of atmospheric turbulence,'' \emph{J.
  Lightw. Technol.}, vol.~27, no.~18, pp. 3965--3973, 2009.

\bibitem{burks1982high}
D.~Burks, E.~Graf, and M.~Fahey, ``A high-frequency analysis of radome-induced
  radar pointing error,'' \emph{IEEE Trans. Antennas Propag.}, vol.~30, no.~5,
  pp. 947--955, 1982.

\bibitem{orzechowski2008optimal}
P.~K. Orzechowski, N.~Y. Chen, J.~S. Gibson, and T.-C. Tsao, ``Optimal
  suppression of laser beam jitter by high-order rls adaptive control,''
  \emph{IEEE Tran. Contr Technol.}, vol.~16, no.~2, pp. 255--267, 2008.

\bibitem{arancibia2005adaptive}
N.~O.~P. Arancibia, N.~Chen, S.~Gibson, and T.-C. Tsao, ``Adaptive control of a
  mems steering mirror for suppression of laser beam jitter,'' in
  \emph{Proceedings of the 2005, American Control Conference, 2005.}\hskip 1em
  plus 0.5em minus 0.4em\relax IEEE, 2005, pp. 3586--3591.

\bibitem{arnon1997performance}
S.~Arnon, S.~R. Rotman, and N.~S. Kopeika, ``Performance limitations of
  free-space optical communication satellite networks due to vibrations:
  direct-detection digital mode,'' in \emph{10th Meeting on Optical Engineering
  in Israel}, vol. 3110.\hskip 1em plus 0.5em minus 0.4em\relax International
  Society for Optics and Photonics, 1997, pp. 357--368.

\bibitem{farid2007outage}
A.~A. Farid and S.~Hranilovic, ``Outage capacity optimization for free-space
  optical links with pointing errors,'' \emph{J. Lightw. Technol.}, vol.~25,
  no.~7, pp. 1702--1710, 2007.

\bibitem{laughlin1995mirror}
M.~J. Laughlin, ``Mirror jitter: an overview of theoretical and experimental
  work,'' \emph{Optical Engineering}, vol.~34, no.~2, pp. 321--330, 1995.

\bibitem{mcever2004adaptive}
M.~A. McEver, D.~G. Cole, and R.~L. Clark, ``Adaptive feedback control of
  optical jitter using q-parameterization,'' \emph{Optical Engineering},
  vol.~43, no.~4, pp. 904--911, 2004.

\bibitem{phillips2003signals}
C.~L. Phillips, J.~M. Parr, and E.~A. Riskin, \emph{Signals, systems, and
  transforms}.\hskip 1em plus 0.5em minus 0.4em\relax Prentice Hall Upper
  Saddle River, 2003.

\bibitem{wang2003simple}
Z.~Wang and G.~B. Giannakis, ``A simple and general parameterization
  quantifying performance in fading channels,'' \emph{IEEE Trans. Commun.},
  vol.~51, no.~8, pp. 1389--1398, 2003.

\bibitem{ko2000outage}
Y.-C. Ko, M.-S. Alouini, and M.~K. Simon, ``Outage probability of diversity
  systems over generalized fading channels,'' \emph{IEEE Trans. Commun.},
  vol.~48, no.~11, pp. 1783--1787, 2000.

\bibitem{yang2014free}
F.~Yang, J.~Cheng, and T.~A. Tsiftsis, ``Free-space optical communication with
  nonzero boresight pointing errors,'' \emph{IEEE Trans. Commun.}, vol.~62,
  no.~2, pp. 713--725, 2014.

\end{thebibliography}
\end{document}